\documentclass[subscriptcorrection,upint,varvw,barcolor=Goldenrod3,mathalfa=cal=euler,balance,hyphenate,french,pdf-a,nolists]{asmejour} %

\hypersetup{%
	pdfauthor={John H. Lienhard},                       		   	
	pdftitle={ASME Journal Paper LaTeX Template},                  	
	pdfkeywords={ASME journal paper, LaTeX template, BibTeX style, asmejour class},
	pdfsubject = {Describes the asmejour LaTeX template},			
	pdflicenseurl={https://ctan.org/pkg/asmejour},
}

\JourName{Dynamic Systems, \\Measurement and Control}

\usepackage{comment}

\newtheorem{proposition}{Proposition}
\newtheorem{remark}{Remark}

\begin{document}
\newcommand{\tbf}[1]{\textbf{#1}}

\newcommand{\mbf}[1]{\mathbf{#1}}
\newcommand{\mbb}[1]{\mathbb{#1}}
\newcommand{\bsym}[1]{\boldsymbol{#1}}
\newcommand{\mcal}[1]{\mathcal{#1}}

\newcommand{\red}[1]{\textcolor{red}{#1}}
\newcommand{\blue}[1]{\textcolor{blue}{#1}}


\SetAuthorBlock{Sriram Narayanan\CorrespondingAuthor, Mrinal Kumar}{Department of Mechanical and Aerospace Engineering\\
The Ohio State University\\
201 W 19th Ave\\
Columbus, 43210, Ohio, USA} 

\SetAuthorBlock{Mohamed Naveed Gul Mohamed, Suman Chakravorty}{Department of Aerospace Engineering\\
Texas A \& M University \\
710 Ross St\\
College Station, 77843, Texas, USA}

\SetAuthorBlock{Indranil Nayak}{Department of Electrical and Computer Engineering\\
The Ohio State University\\
2015 Neil Ave\\
Columbus, 43210, Ohio, USA} 


\title{On the Predictive Capability of Dynamic Mode Decomposition for Nonlinear Periodic Systems with Focus on Orbital Mechanics}


\begin{abstract}
This paper discusses the predictive capability of Dynamic Mode Decomposition (DMD) in the context of orbital mechanics. The focus is specifically on the Hankel variant of DMD which uses a stacked set of time-delayed observations for system identification and subsequent prediction. A theory on the minimum number of time delays required for accurate reconstruction of periodic trajectories of nonlinear systems is presented and corroborated using experimental analysis. In addition, the window size for training and prediction regions, respectively, is presented. 
The need for a meticulous approach while using DMD is emphasized by drawing comparisons between its performance on two candidate satellites, the ISS and MOLNIYA-3-50. 
\end{abstract}

\date{Version \versionno, \today}

\maketitle 


\section{Introduction}
This article considers the predictive capability of Dynamic Mode Decomposition (DMD) in modeling nonlinear periodic systems, with emphasis on its application in orbital mechanics. DMD was first introduced in the field of fluid mechanics to develop reduced order models, but has since seen applications in many other fields~\citep{schmid2010dynamic, schmid2022dynamic, kutz2016dynamic}. DMD offers a viable alternative to conventional analytical models that satellite operators currently depend on, by providing reasonably accurate predictions of satellite state vectors from historical time series data. Data-driven machine learning techniques have become increasingly popular in the orbit determination and propagation community in recent years. Researchers have demonstrated the superior abstraction capabilities of such methods through increased prediction accuracy and ease of implementation. Several methods, including distributed regression, transfer learning ~\cite{sharma2015robust}, Gaussian mixture models ~\citep{terejanu2008uncertainty, demars2013entropy}, feed-forward neural networks ~\cite{peng2019comparative}, support vectors machines ~\cite{peng2018exploring}, Gaussian processes ~\cite{peng2019gaussian}, convolutional neural networks ~\cite{pihlajasalo2018improvement}, and Long short-term memory (LSTMs) networks have been examined for orbit prediction ~\cite{azmi2021adaptation}. Recent work has demonstrated DMD and delay embedding to construct a data-driven real-time guidance and control system for asteroid landings based on the restricted three-body problem ~\cite{kajikawa2023data}. In this work, the authors build upon DMD with control, which involves devising appropriate control laws once the system's state matrices are obtained ~\cite{proctor2016dynamic}. The approach in Ref.~\cite{kajikawa2023data}  differs from the present article in that a delay embedding-based DMD is applied to observations of dynamics based on the two-body problem. Also, there is work that shows the application of Koopman operator theory to generate analytical solutions for satellite orbits affected by zonal harmonics ~\cite{arnas2021approximate}. The Koopman operator has also been used to reformulate nonlinear optimal control problems in space flight through bilinearization and the Galerkin method to model the time evolution as a linear combination of appropriate basis functions~\cite{hofmann2022advances}. 
Further, closed-form solutions of the Koopman operator have been calculated using the Galerkin approximation for astrodynamics applications~\cite{linares2019koopman}. The authors in Ref.~\cite{linares2019koopman}, also present a comparison of their methods against a purely data-driven approach using an extended version of DMD and a Fourier-Hermite basis. For attitude control, Koopman operator theory has been used to develop control laws for higher dimensional linear systems based on observables that portray nonlinear attitude dynamics on SO(3) ~\cite{Chen2023}. Koopman operator theory has also been applied to generate robust control laws for control about an equilibrium point in the circular restricted three-body problem (CR3BP) with control-dependent noise system~\cite{jensonrobust}. There is also work that connects the Koopman operator to state-space-based systems identification methods such as the eigen realization algorithm (ERA) ~\citep{proctor2016dynamic, juang1988eigensystem}. It is important to note that the Hankel matrix formulation in both ERA and DMD are similar and the subspace system matrix generated by the two methods is related via a similarity transform ~\cite{tu2013dynamic}. This approach has further been extended to time-varying models as well ~\citep{gueho2021time,majji2010time}. Comparisons between DMD and autoregressive moving average (ARMA) models have been drawn in several instances and their similarity is well-known in the literature. It is noted that  DMD makes forecasts by connecting each time snapshot directly to the previous one, in a manner very similar to ARMA models. Variations such as VARIMA (vector ARIMA) and SARIMA (seasonal ARIMA) exist to accommodate multiple time series and seasonal effects. In the DMD framework, seasonal fluctuations are inherently considered. DMD takes advantage of the data vectors by capturing both oscillatory patterns and exponential trends via its spectrum~\cite{brunton2021modern}. The use of ARMA models for the identification of time-varying linear systems is also addressed in the literature ~\cite{mohamedinformation}. 
Additionally, it has been shown that combining machine learning models with sequential estimation techniques improves simultaneous tracking and navigation (STAN) for orbit determination. ML models are shown to fuse observations, handle nonlinearities, augment the state vector, and adapt estimation parameters. Sequential estimation algorithms, e.g., the Kalman filter, utilize combined observations from several sensors and incorporate ML outputs for accurate tracking and uncertainty propagation. This synergy optimizes the orbit determination procedure by leveraging ML's capabilities and the recursive nature of sequential estimation, thereby improving its accuracy ~\citep{mortlock2021assessing, kozhaya2021comparison}. Typically, statistical orbit determination and prediction methods are used to learn trajectories from observations. They include the Kalman Filter, Extended Kalman Filter (EKF), Unscented Kalman Filter (UKF), and Particle Filter. They utilize statistical techniques to estimate and predict satellite trajectories while accounting for observation uncertainties. These methods vary in their approaches, from iterative estimation with Gaussian error distributions (Kalman Filter) to handling nonlinearity (EKF and UKF), and non-Gaussian systems (Particle Filter). Here the choice of method depends on factors like specific application, available data, and desired accuracy and computational complexity. All methods described so far depend on the availability of an approximate analytical model of the dynamics. The Kalman Filter is not suitable for nonlinear systems or non-Gaussian noise. The EKF can introduce errors due to local linearization, and the UKF is limited to Gaussian closure. The Particle Filter is computationally demanding, especially for high-dimensional systems, and can encounter particle degeneracy  ~\citep{schutz2004statistical, crassidis2004optimal}. The alternative of using a purely data-driven, ``equation-free'' method is shown here with specific situations where it can be applied. The distinction between filtering and data-driven estimation is quite evident, filtering is a two-step process that involves estimating, forecasting the variable of interest, and updating the initial model based on which forecasts are made. This work only deals with data-driven model estimation and trajectory forecasting, whereby comparisons are drawn purely in that respect. It is important to note that this paper focuses solely on using DMD for the prediction of trajectories and not on performing a measurement update in the sense of recursive state estimation (Bayesian filtering). 
The contributions of this paper are as follows:
\begin{itemize}
    \item Demonstration of a DMD-based approach for learning and predicting satellite state vectors from historical trajectory observations, offering insight into DMD's predictive capabilities in this context. Theoretical reasoning is provided for why periodic trajectories of nonlinear systems can be modeled as a linear autoregressive model and specifically using the DMD algorithm as a data-driven method in this work. A lower bound for the minimal number of time delays required to model a nonlinear periodic trajectory is provided. 

    \item Discussion on selection of suitable DMD model parameters such as the sampling interval, training window size, and the prediction window can be identified in the context of trajectories in Earth orbit. 
\end{itemize}
Analysis performed on benchmark dynamical systems such as the simple pendulum and curated model datasets representing various perturbations encountered by candidate satellites is presented. DMD is employed to generate data-driven surrogate dynamic models for two candidate satellites, the International Space Station (ISS-ZARYA) and Molniya-3-50. This selection allows a comparison of results between a nearly circular orbit (ISS) and a highly eccentric orbit (Molniya). This study starts with a basic Keplerian model free of perturbations and goes up to the Simplified General Perturbations model (SGP4), currently used to propagate two line elements (TLE). In the Molniya-3-50 case, for conciseness, analysis for the scenario that includes all perturbations is only presented. The corresponding dataset was generated using the deep space variant of SGP4, the Simplified Deep Perturbations model (SDP4).


This paper is organized as follows. Sec.~\ref{Sec:PF} formulates the problem considered and defines the objective of this work. Sec.~\ref{Sec:Th} discusses the theoretical background of using AR models to predict the evolution of periodic orbits of nonlinear systems using a finite number of harmonics. Sec.~\ref{Sec:Algo} discusses the DMD algorithm, its relation to the Koopman operator, and the time-delayed variant, called Hankel-DMD. Sec.~\ref{Sec:Res} presents observations and results from the experimental analysis. 
Finally, Sec.~\ref{Sec:Conc} provides a general discussion summarizing the main contributions of this work and concludes the paper.


\section{Problem Formulation}
\label{Sec:PF}
Consider a nonlinear system
\begin{equation}
    \mbf{x}(k+1) = \mbf{f}(\mbf{x}(k)) 
    \label{Eq:state_equation}
\end{equation}
with known initial condition $\mbf{x}(0)$, where $\mbf{x}(k) \in \mbb{R}^n$. Assume that the system has a non-constant periodic solution i.e. $\exists \ N > 0$ such that $\mbf{x}(k+N) = \mbf{x}(k) \ \forall \ k$. The dynamics $\mbf{f}(\cdot)$ of the system is unknown. 
Given, a finite time history of the states $\{\mbf{x}(k), \mbf{x}(k+1), \cdots, \mbf{x}(k+l) \}$, the objective is to find a surrogate model for this nonlinear system on its periodic orbit using the DMD algorithm. The surrogate model will then be used to predict the next state $\mbf{x}(k+l+1)$, and subsequently, forecast the evolution of the system. In addition, we seek to identify the number of time delays required, the size of the training window, and how far the predictions can be made using these two model parameters. In Sec.~\ref{Sec:Res}, the findings are illustrated using examples from the  orbital mechanics domain. 

\red{}

\section{Theoretical Background}
\label{Sec:Th}


This section describes the background behind modeling a periodic trajectory of the system in Eq.~\eqref{Eq:state_equation} as a linear autoregressive (AR) model and subsequently using the DMD algorithm described in Sec.~\eqref{Sec:Algo} for modeling the system. The analysis presented below is similar to the work done in Ref.~\cite{pan2020structure}, but it is an alternate concise development and shows that there always exists an AR model for a periodic orbit (of a nonlinear system) with a finite number of harmonics. Further, the minimal time delays required can be construed from the data matrix (and can be related to the number of dominant frequencies in the orbit).

Consider the system shown in Eq.~\eqref{Eq:state_equation} and the associated periodic solution $\mbf{x}(k)$. Since $\mbf{x}(k)$ is a discrete-time signal with period $N$, its Fourier series can be written as follows \cite{signals_systems}, 

\begin{align}
    \mbf{x}(k) = \sum_{m = 0}^{N-1} \mbf{a}_m e^{j \omega_m k}, \label{Eq:Fourier_series}
\end{align}
where, $\mbf{a}_m \in \mathbb{C}^n$ are Fourier coefficients and $\phi_m(k) = e^{j \omega_m k}$ are the basis functions. The coefficients $\mbf{a}_m$ can be written in terms of the signal $\mbf{x}(k)$ as 
\begin{align}
    \mbf{a}_m = \sum_{k=0}^{N-1} \mbf{x}(k)e^{-j \omega_m k}.
\end{align}
Note that the period, $N$ may vary depending on the initial condition $x(0)$ of the system in Eq.~\eqref{Eq:state_equation}. Here, we model only a single trajectory from $\mathbf{x}(0)$, whose period is $N$.

For the sake of simplicity, we assume the signal does not have a constant bias term, and hence $N$ is even for real signals. Due to the Nyquist criterion, the Fourier series in Eq.~\eqref{Eq:Fourier_series} can have a maximum of $N/2$ frequencies. 
Since both the negative and the positive components are represented, Eq.~\eqref{Eq:Fourier_series} has $2\times (N/2)$ terms. If the signal $\mbf{x}(k)$ has only $M$ frequencies, where, $1 \leq M \leq N/2 $, then, the Fourier series can be written as 
\begin{align}
    \mbf{x}(k) = \sum_{m = 0}^{2M-1} \mbf{a}_m e^{j \omega_m k}, \label{Eq:Fourier_series_M}
\end{align}
where, $\{\omega_0, \omega_1, \cdots, \omega_{2M-1} \}$ are the different frequencies in the signal. 

Given the system's response $\{\mbf{x}(k), \mbf{x}(k+1), \cdots \mbf{x}(k+l)\}$, one can estimate the Fourier coefficients by solving a linear least squares problem as shown below.
The signal samples at $l$ consecutive time-steps can be written as:

\begin{align}
     \mbf{x}(k) &= \mbf{a}_0 e^{j \omega_0 k} + \mbf{a}_1 e^{j \omega_1 k} + \cdots + \mbf{a}_{2M-1} e^{j \omega_{2M-1} k}, \label{Eq:x_exp}\\
      \mbf{x}(k+1) &= \mbf{a}_0 e^{j \omega_0 (k+1)} + \mbf{a}_1 e^{j \omega_1 (k+1)} + \cdots  \nonumber \\ &+ \quad \mbf{a}_{2M-1} e^{j \omega_{2M-1} (k+1)}, \\
      & \vdots  \nonumber\\
      \mbf{x}(k+l-1) &= \mbf{a}_0 e^{j \omega_0 (k+l-1)} + \mbf{a}_1 e^{j \omega_1 (k+l-1)} + \cdots \nonumber\\
      &+ \quad \mbf{a}_{2M-1} e^{j \omega_{2M-1} (k+l-1)}\label{Eq:x_exp1}.  
\end{align}
Eqs.~\eqref{Eq:x_exp}-\eqref{Eq:x_exp1} can be written as:
\begin{align}
    [\mbf{x}(k), \mbf{x}(k+1), \cdots, \mbf{x}(k+l-1)] = [\mbf{a}_0, \mbf{a}_1, \cdots, \mbf{a}_{2M-1}] \mbf{V},
\end{align}
where,
\begin{equation}
    \mbf{V} = 
    \begin{bmatrix}
        e^{j \omega_0 k} & e^{j \omega_0 (k+1)} & \cdots & e^{j \omega_0 (k+l-1)} \\
        e^{j \omega_1 k} & e^{j \omega_1 (k+1)} &\cdots & e^{j \omega_1 (k+l-1)}\\
        \vdots & \vdots &\ddots & \vdots  \\
        e^{j \omega_{2M-1} k} &e^{j \omega_{2M-1} (k+1)}& \cdots & e^{j \omega_{2M-1} (k+l-1)}
    \end{bmatrix}.
    \label{Eq:Vandermonde_mat}
\end{equation}
$\mbf{V} \in \mathbb{C}^{2M \times l}$ and is a Vandermonde matrix and has linearly independent rows. To calculate the Fourier coefficients, the $rank(\mbf{V}) = 2M$, hence the number of samples $l \geq 2M$. 
\begin{align}
    [\mbf{a}_0, \mbf{a}_1, \cdots, \mbf{a}_{2M-1}] &= [\mbf{x}(k), \mbf{x}(k+1), \cdots, \mbf{x}(k+l-1)] \mbf{V}^{\dagger}, \label{Eq:LS_sol}\\
    \text{where~}\mbf{V}^{\dagger} &= \mbf{V}^{\mathsf{T}} (\mbf{V}\mbf{V}^{\mathsf{T}})^{-1}. \label{Eq:pinv}     
\end{align}

$\mbf{x}(k+l) $ can be written as,

\begin{align}
    \mbf{x}(k+l) =  
    \begin{bmatrix}
        \mbf{a}_0 & \mbf{a}_1 & \cdots & \mbf{a}_{2M-1}    
    \end{bmatrix} 
    \begin{bmatrix}
        e^{j \omega_0 (k+l)} \\
        e^{j \omega_1 (k+l)} \\
        \vdots \\
        e^{j \omega_{2M-1} (k+l)}
    \end{bmatrix}. \label{Eq:prediction}
\end{align}
Substituting Eq.~\eqref{Eq:LS_sol} in Eq.~\eqref{Eq:prediction} gives,
\begin{align}
    \mbf{x}(k+l) &= [\mbf{x}(k), \mbf{x}(k+1), \cdots, \mbf{x}(k+l-1)]\mbf{V}^{\dagger}
     \underbrace{\begin{bmatrix}
        e^{j \omega_0 (k+l)} \\
        e^{j \omega_1 (k+l)} \\
        \vdots \\
        e^{j \omega_{2M-1} (k+l)}
    \end{bmatrix}}_{\mbf{Y}}.
\end{align}
$\mbf{V}$ and $\mbf{Y}$ can also be written as,
\begin{align*}
    \mbf{V} =& 
    \begin{bmatrix}
        e^{j \omega_0} & 0 & 0 & \cdots & 0 \\
        0 & e^{j \omega_1} & 0 & \cdots & 0 \\
        \vdots & \vdots & & \ddots   & \vdots \\
        0 & 0 & 0 &\cdots  &e^{j \omega_{2M-1}}
    \end{bmatrix}^k
    \times \\
    & \quad \begin{bmatrix}
        1 & e^{j \omega_0} & \cdots & e^{j \omega_0 (l-1)}\\
        1 & e^{j \omega_1} & \cdots & e^{j \omega_1 (l-1)}\\
        \vdots & \vdots & \ddots & \vdots \\
        1 & e^{j \omega_{2M-1}} & \cdots & e^{j \omega_{2M-1} (l-1)}
    \end{bmatrix}, \\
    \mbf{Y} = & 
    \begin{bmatrix}
        e^{j \omega_0} & 0 & 0 & \cdots & 0 \\
        0 & e^{j \omega_1} & 0 & \cdots & 0 \\
        \vdots & \vdots & & \ddots   & \vdots \\
        0 & 0 & 0 &\cdots  &e^{j \omega_{2M-1}}
    \end{bmatrix}^k
    \begin{bmatrix}
        e^{j \omega_0 l} \\
        e^{j \omega_1 l} \\
        \vdots \\
        e^{j \omega_{2M-1} l} \\
    \end{bmatrix}.
\end{align*}
Then,
\begin{align}
    \bar{\boldsymbol{\alpha}} &:= \mbf{V}^{\dagger} \mbf{Y}, \nonumber\\
                  &=  \begin{bmatrix}
        1 & e^{j \omega_0} & \cdots & e^{j \omega_0 (l-1)}\\
        1 & e^{j \omega_1} & \cdots & e^{j \omega_1 (l-1)}\\
        \vdots & \vdots & \ddots & \vdots \\
        1 & e^{j \omega_{2M-1}} & \cdots & e^{j \omega_{2M-1} (l-1)}
    \end{bmatrix}^{\dagger}
    \begin{bmatrix}
        e^{j \omega_0 l} \\
        e^{j \omega_1 l} \\
        \vdots \\
        e^{j \omega_{2M-1} l} \\
    \end{bmatrix} \label{Eq:AR_model_parameters}
\end{align}
where, $\bar{\boldsymbol{\alpha}} = [\alpha_0, \alpha_1, \dots, \alpha_{l-1}]^{\mathsf{T}}$. Note that $\bar{\boldsymbol{\alpha}}$ is independent of the time $k$, and hence we obtain a time-invariant autoregressive model for the trajectory of the system in Eq.~\eqref{Eq:state_equation}: 
\begin{align}
    \mbf{x}(k+l) & = \alpha_0 \mbf{x}(k) + \alpha_1 \mbf{x}(k+1) + \cdots +\alpha_{l-1} \mbf{x}(k+l-1) \label{Eq:AR_model}.
\end{align}
Eq.~\eqref{Eq:AR_model} shows that the response of the system in Eq.~\eqref{Eq:state_equation} can be written as a linear autoregressive (AR) model using a finite history of states from the past. Due to the rank condition of the matrix $\mbf{V}$, the length of the history $l$ or the number of time delays should be such that $l\geq 2M$, where $M$ is the number of modes/frequencies in the signal. We summarize the above result. 

\begin{proposition}\label{prop:1}
    Given a nonlinear dynamical system as in Eq.~\eqref{Eq:state_equation}, a periodic trajectory from the system can always be modeled using a linear autoregressive model as shown in Eq.~\eqref{Eq:AR_model}. The minimal number of time delays required for the AR model is twice the number of frequencies $M$  present in the signal $\mbf{x}(k)$. 
\end{proposition}
\textit{Proof:} From Lemma 1 in Ref.~\cite{pan2020structure}, the Vandermonde matrix $\mbf{V}$ has rank $2M$, given $l \geq 2M$. It can be seen from the development in Eqs.~\eqref{Eq:LS_sol} - \eqref{Eq:AR_model_parameters} that there always exists a solution to the AR model parameters which model the periodic trajectory using Eq.~\eqref{Eq:AR_model}. 
\hfill $\blacksquare$

In practice, one does not have access to the fundamental frequency and the number of harmonics of the signal to estimate the AR model parameters as shown in Eq.~\eqref{Eq:AR_model_parameters}. The next section discusses how one can construct the AR model and find the minimal number of time-delays directly from the observed data.     
\section{Methodology: Time - Delay Dynamic Mode Decomposition or Hankel-DMD}
\label{Sec:Algo}
Considering the discrete-time dynamic system in Eq.~\eqref{Eq:state_equation}, the discrete-time Koopman operator $\mathcal{K}$ operates on the space of observables, $g(\mbf{x})$ such that, 
\begin{equation}
 \mathcal{K} g\left(\mbf{x}(k)\right) \triangleq \left(g \circ \mbf{f}\right)\left(\mbf{x}(k)\right)=g\left(\mbf{x}(k+1)\right)
 \label{Eq:koopman}
\end{equation}
The Koopman operator $\mathcal{K}$ acts linearly on the observables $g$ even though the dynamics defined by $\mbf{f}$ is nonlinear. The eigendecomposition of the Koopman operator contains point, continuous, and residual spectra with a generalized eigendistribution ~\cite{mohr2014construction}. This work only considers the point spectra, as seen in Eq.~\eqref{eq:koop1},
\begin{align}
    \mathcal{K}\theta_m(\mbf{x}) = \lambda_m \theta_m(\mbf{x}), \ \ m = 0,1,2,....,N-1
    \label{eq:koop1}
\end{align}
where $\theta_m(\mbf{x})$ and $\lambda_m$ are corresponding Koopman eigenfunctions and eigenvalues. Assuming eigenfunctions span the observable space, a vector-valued observable $\mbf{h}$ is defined in terms of the eigenfunctions,
\begin{align}
    \mbf{h}(\mbf{x}) =  \sum_{m=0}^{\infty} \theta_m (\mbf{x})\bsym{\psi}_m, \ \  \bsym{\psi}_m \in \mathbb{C}^n
    \label{eq:koop2}
\end{align}
Substituting Eq.~\eqref{eq:koop1} into Eq.~\eqref{eq:koop2}~~\cite{nayak2021detection}, 
\begin{align}
    \mbf{h}(\mbf{F}^k(\mbf{x}(0))) = 
    \mbf{h}(\mbf{x}(k)) =  \sum_{m=0}^{\infty} \lambda_m^k \theta_m (\mbf{x}(0))\bsym{\psi}_m, \ \  \bsym{\psi}_m \in \mathbb{C}^n
    \label{eq:koop3}
\end{align}
where $\bsym{\psi}_m$ are the Koopman modes. If the phase space is restricted to periodic trajectories, the Koopman modes are given by the discrete Fourier transform (DFT) \cite{rowley2009spectral}. From Eq.~\eqref{Eq:Fourier_series_M},
\begin{align}
    \mbf{x}(k) &= \sum_{m = 0}^{2M-1} \mbf{a}_m e^{j \omega_m k}
    = \sum_{m = 0}^{2M-1} \mbf{a}_m \theta_m(\mbf{x}(k))
\end{align}
where $\theta_m(\mbf{x})$ are the eigenfunctions of the Koopman operator $\mathcal{K}$, with eigenvalues $\lambda_m = e^{j \omega_m}$. Eq.~\eqref{eq:koop3} can be rewritten as,
\begin{align}
    \mbf{h}(\mbf{x}(k)) &=  \sum_{m=0}^{2M-1} \lambda_m^k \theta_m (\mbf{x}(0))\mbf{\hat{a}_m}, \ \  \mbf{a}_m \in \mathbb{C}^n 
\end{align}
where the Koopman modes $\bsym{\psi}_m$ are given by the Fourier coefficients $\mbf{{a}}_m$. The time expansion of vector-valued observable $\mbf{h}$ completely determines the evolution of the nonlinear periodic system from its initial condition. However, it must be noted that there is no explicit representation of the Koopman operator. Instead, the operator's behavior is found through its action on a finite-dimensional observable space. Data-driven techniques, like DMD, are employed to approximate the Koopman modes. DMD aims to find the most accurate approximation of $\mathcal{K}$ within a finite-dimensional space. The spatiotemporal modes derived through DMD exhibit convergence to the Koopman modes when applied to a specific set of linear observables while assuming that state variables themselves serve as the set of observables $\mbf{h(x)} = \mbf{x}$ ~\citep{rowley2009spectral,tu2013dynamic}.  DMD operates on a pair of snapshot matrices to generate a linear model approximating the nonlinear periodic trajectory. The snapshot matrices are created by sampling the state vectors of dimension $(n \times 1)$ at different time $(k)$ instants to form snapshot matrices $\mbf{X}_k$ and $\mbf{X}_{k+1}$, where $\mbf{X}_{k+1}$ is one-time snapshot ahead of the original snapshot matrix $\mbf{X}_k$.
\begin{align}
    \mbf{X}_k &= 
    \begin{bmatrix}
        \mbf{x}(k) & \mbf{x}(k+1) & \mbf{x}(k+2) & ... & ... & ... & \mbf{x}(k+s-1)\\
    \end{bmatrix}\label{Eq:state_matrix}
\end{align}
\begin{align}
    \mbf{X}_{k+1} &= 
    \begin{bmatrix}\label{eq:dmd_snap_mat}
        \mbf{x}(k+1) & \mbf{x}(k+2) & \mbf{x}(k+3) & ... & ... & ... & \mbf{x}(k+s)\\
    \end{bmatrix}
\end{align}
Eq.~\eqref{eq:dmd_snap_mat} characterizes the time instants denoted as $k$ and the time step $\Delta k$ between consecutive snapshots. Without loss of generality, it is assumed that the sampling process is uniform, and the snapshots commence at $k=0$. A total of $s$ snapshots are collected inside the DMD training window. Here, it is assumed that $s >> l$ i.e. the training window is much larger than the time delay dimension required. This additional temporal length is used to improve spatial resolution using the Hankel-DMD algorithm, as explained below.

The DMD algorithm identifies an optimal linear operator denoted as ${\mbf{A}}$ that establishes the relationship between the two snapshot matrices, $\mbf{X}_k$ and $\mbf{X}_{k+1}$, in a least-square sense, as seen in Eq.~\eqref{eq:dmd_linear_fit}. The ${\mbf{A}}$ matrix represents the Koopman operator in the newly mapped linear space of finite-dimensional observables.
\begin{align}\label{eq:dmd_linear_fit}
    \mbf{X}_{k+1} &\approx {\mbf{A}\mbf{X}_k} 
\end{align}
it follows that,
\begin{align}
    \mbf{X}_{k+l} &\approx {\mbf{A}^l\mbf{X}_k} 
\end{align}
From Eq.~\eqref{Eq:AR_model},
\begin{align}
    \mbf{x}(k+l) & = \alpha_0 \mbf{x}(k) + \alpha_1 \mbf{x}(k+1) + \cdots +\alpha_{l-1} \mbf{x}(k+l-1)
\end{align}
Consider $\mbf{L}$ such that,
\begin{align}
    \mbf{L} &= 
    \begin{bmatrix}
         \mbf{x}(k) & \mbf{A}\mbf{x}(k) & \mbf{A}^2\mbf{x}(k) & ... & ... & ... & \mbf{A}^{l-1}\mbf{x}(k+l-1) 
    \end{bmatrix} \nonumber \\
    &= 
    \begin{bmatrix}
         \mbf{x}(k) & \mbf{x}(k+1) & \mbf{x}(k+2) & ... & ... & ... & \mbf{x}(k+l-1)\\
    \end{bmatrix}
\end{align}
then it is seen that,
\begin{align}
    \mbf{AL} = \mbf{LC} \label{Eq:comp_eigen}
\end{align}
where $\mbf{C}$ is a companion matrix,
\begin{align}
    \mbf{C} &=
    \begin{bmatrix}
        0 &  0 & \cdots & 0 &\alpha_0 \\
        1 &  0 & \cdots & 0 &\alpha_1 \\
        \vdots & \vdots & \ddots & \vdots & \vdots\\
        0 & \cdots & \cdots & 1 & \alpha_{l-1}
    \end{bmatrix}
\end{align}
$\mbf{C}$ is related to the Vandermonde matrix $\mbf{V}$ from Eq.~\eqref{Eq:Vandermonde_mat} in the following manner, $\mbf{C} = \mbf{V^{\dagger} D V}$, where $\mbf{D} = diag(e^{j \omega_0}, e^{j \omega_1}, \cdots, e^{j \omega_{2M-1}})$ and $\dagger$ is the Moore-Penrose inverse.
The eigenvalues of $\mbf{C}$ are the subset of the eigenvalues of $\mbf{A}$ if,
\begin{align}
    \mbf{Ca} = \lambda \mbf{a}
\end{align}
and using Eq.~\eqref{Eq:comp_eigen}, it can be verified $\mbf{v} = \mbf{La}$ is an eigenvector of $\mbf{A}$ with eigenvalue $\lambda$ ~\cite{rowley2009spectral}. Here, the assumption of $\mbf{X}_k$ being full rank is generally not true in practice 
as the data matrices $\mbf{X}_k,\mbf{X}_{k+1}$ are typically rank deficient. This is especially true when the data stems from experiments contaminated with noise and other uncertainties. SVD-based DMD offers a more robust approach by operating in a lower dimensional space spanned by the dominant modes of the dataset. These modes are often called the proper orthogonal modes (POD). To find the reduced order space, the singular value decomposition (SVD) of snapshot matrix $\mbf{X}_k$ is computed. 
\begin{align}
    \mbf{X}_k &= \mbf{U} \mbf{\Sigma} \mbf{V}^{H} \approx \bsym{\Tilde{U}} \bsym{\Tilde{\Sigma}} \bsym{\Tilde{V}}^{H},
\end{align}
where the superscript $H$ denotes the Hermitian transpose. The matrix $\mbf{U}$ contains the left singular vectors, $\mbf{\Sigma}$ is a diagonal matrix of singular values arranged hierarchically and the matrix $\mbf{V}$ contains the right singular vectors of $\mbf{X}_k$. In practice, it is sufficient to compute the reduced SVD, where only the first few $(r)$ columns and rows of respectively $\mbf{U}$ and $\mbf{V}$ are retained with the first $r$ singular values of $\mbf{\Sigma}$. The SVD energy thresholding algorithm is used to determine $r$ and then truncate our $\mbf{U}$, $\mbf{\Sigma}$, and $\mbf{V}$ matrices accordingly ~\cite{kutz2013data}. Next, the matrix $\bsym{\Tilde{A}}$, a projection of $\mbf{A}$ in the lower dimensional space is calculated.
\begin{align}
    {\mbf{A}} &= \mbf{X}_{k+1} \mbf{X}_k^{\dagger}
     = \mbf{X}_{k+1} \bsym{\Tilde{V}}  \bsym{\Tilde{\Sigma}}^{-1 } \bsym{\Tilde{U}}^{H} 
\end{align}
 ${\mbf{A}}$ is then projected onto a lower dimensional space using a similarity transformation
\begin{align}
    {\bsym{\Tilde{A}}} &= \bsym{\Tilde{U}}^{H} {\mbf{A}} \bsym{\Tilde{U}} 
 = \bsym{\Tilde{U}}^{H} \mbf{X}_{k+1} \bsym{\Tilde{V}} \bsym{\Tilde{\Sigma}}^{-1 }
\end{align}
It should be noted that the eigenvalues of ${\mbf{A}}$ are adequately approximated by the eigenvalues of ${\bsym{\Tilde{A}}}$. The eigendecomposition of $\bsym{\Tilde{A}}$ is performed to extract the global modes present in the dataset. This process is akin to the standard Arnoldi method ~\citep{schmid2010dynamic,rowley2009spectral}. 
\begin{align}
    {\bsym{\Tilde{A}}} \mbf{W}  &= \mbf{W} \mbf{\Lambda}  \label{eq:Atilde}
\end{align}
$\mbf{W}$ is the matrix of eigenvectors of ${\bsym{\Tilde{A}}}$ and $\mbf{\Lambda}$ is the diagonal matrix containing its eigenvalues. The DMD modes are obtained and given by the columns of $\mbf{Z} = \mbf{X}_{k+1} \bsym{\Tilde{V}} \bsym{\Tilde{\Sigma}}^{-1} \mbf{W}$. These DMD modes approximate the eigenvectors of the high dimensional matrix ${\mbf{A}}$ ~\citep{tu2013dynamic, kutz2016dynamic}. Note that this formulation follows \textit{exact}-DMD and differs from the original formulation which defines $\mbf{Z} = \bsym{\Tilde{U}W}$ as the projected DMD modes. Finally, the DMD predicted state $\mbf{\hat{x}}(k)$ at any time $k$ is obtained by
\begin{align}
    \bsym{\hat{x}}(k) &= \mbf{Z} \exp(\mbf{\Omega} k) \mbf{b} 
    \label{eq:DMD_time_dynamics}
\end{align}
where $ \mbf{b} = \mbf{Z}^{\dagger} \mbf{x}(0)$, is the scaling factor or magnitude of DMD modes. In the above developments, $\mbf{x}(0)$ is the first column of $\mbf{X}$ and, $\mbf{\Omega}$ is the vector of DMD frequencies ($\omega$) which are obtained from DMD eigenvalues using the relation $\omega=\frac{\ln\lambda}{\Delta k}$, with $\lambda$ being the diagonal elements of $\mbf{\Lambda}$~\cite{brunton2022data}. The following sub-section covers the Hankel variant of the DMD algorithm, specifically used in this work to allow time delay embedding in the snapshot matrices. Thus improving the spatial resolution and meeting the time delays requirement to construct a periodic trajectory with $M$ frequencies. This technique is especially useful when $n << M$ i.e. the state dimensions are much smaller than the number of frequencies in the signal.

\subsection{Hankel-DMD}
Hankel reconstruction or time delay embedding is the process of arranging state observations recursively to increase the spatial dimensions of the dataset. 
Delay embedding has been shown to represent ergodic attractors in nonlinear dynamical systems ~\cite{takens1981detecting}. This method finds applications in various domains, including signal analysis and forecasting~\citep{box2015time,brunton2017chaos,arbabi2017ergodic,le2017higher}. Considering the state matrix in Eq.~\eqref{Eq:state_matrix}, the Hankel matrix $\mbf{H}_k$ can be constructed as,
\begin{align}
\mbf{H}_k &= 
    \left[\begin{array}{c}
\mathbf{X}_{k} \\
\mathbf{X}_{k+1} \\
\vdots \\
\mathbf{X}_{k+l-2} \\
\mathbf{X}_{k+l-1}
\end{array}\right] \in \mathbb{R}^{nl \times (s)}, \label{Eq:data_matrix}
\end{align}
where $l$ is the delay embedding dimension. Delay embedding is also vital when the spectral complexity of a dynamical system surpasses its spatial complexity. This scenario commonly occurs in systems with a wide spectrum of insufficient spatial sampling coordinates. Instead of a tall-and-skinny matrix, the data matrix is short and wide, and the DMD algorithm fails to extract all pertinent spectral features. To tackle this problem, a higher-order Dynamic Mode Decomposition (DMD) method is used ~\cite{le2017higher}. It relies on a delay embedding across multiple dimensions, with,
\begin{equation}
\mathbf{X}_{k+l} \approx \mathbf{A}_0 \mathbf{X}_{k}+\mathbf{A}_1 \mathbf{X}_{k+1}+\ldots+\mathbf{A}_{l-1} \mathbf{X}_{k+l-1},
\end{equation}
where note that this embedding is precisely the AR model developed in the previous section which we have already shown can be used to model a periodic orbit of a nonlinear system.
The resulting first-order problem can be written as ~\cite{nayak2023fly},
\begin{equation}
    \mbf{H}_{k+1} = \mbf{C}_{H_k}^{H_{k+1}} \mbf{H}_{k}
\end{equation}
\begin{equation}
\left[\begin{array}{c}
\mathbf{X}_{k+1} \\
\mathbf{X}_{k+2} \\
\vdots \\
\mathbf{X}_{k+l-1} \\
\mathbf{X}_{k+l}
\end{array}\right] \approx\left[\begin{array}{ccccc}
\mathbf{0}_n & \mathbf{I}_n & \ldots & \mathbf{0}_n & \mathbf{0}_n \\
\mathbf{0}_n & \mathbf{0}_n & \ldots & \mathbf{0}_n & \mathbf{0}_n \\
\vdots & \vdots & \ddots & \vdots & \vdots \\
\mathbf{0}_n & \mathbf{0}_n & \ldots & \mathbf{0}_n & \mathbf{I}_n \\
\mathbf{A}_0 & \mathbf{A}_{1} & \ldots & \mathbf{A}_{l-2} & \mathbf{A}_{l-1}
\end{array}\right]\left[\begin{array}{c}
\mathbf{X}_{k} \\
\mathbf{X}_{k+1} \\
\vdots \\
\mathbf{X}_{k+l-2} \\
\mathbf{X}_{k+l-1}
\end{array}\right]
\end{equation}
where $\mbf{0}_n$ and $\mbf{I}_n$ are zero and identity matrices of dimension $n \times n$. And $\mbf{C}_{H_k}^{H_{k+1}}$ is the block matrix that relates the two Hankel matrices $\mbf{H}_k$ and $\mbf{H}_{k+1}$. For cases when $\mbf{X}_{k+1} = \mbf{AX}_k$ exactly, then $\mbf{A}_0 = \mbf{A}, \mbf{A}_1 = \mbf{A}^2, \cdots, \mbf{A}_{l-1} = \mbf{A}^l$. Both Hankel-DMD and regular DMD produce the same set of eigenvalues and modes after truncation \cite{nayak2023fly}.
The higher-order extension adds resilience and adaptability to the conventional algorithm, allowing analysis of systems where temporal length can be traded to improve spatial resolution. This work uses Hankel-DMD for all analysis. The resulting change is that $\mbf{H}_k$ and $\mbf{H}_{k+1}$ are used as inputs to the DMD algorithm. We can also directly estimate the AR model but prefer to use the Hankel-DMD form above as it is more flexible.
The advantage of this algorithm is that one can find the minimal number of time delays required for the model directly from the data. First, we guess the value of $l$ and build the data matrix $\mbf{H}_k$ such that the number of columns is greater than the number of rows, i.e. $s > n l$. The minimum number of time delays is given by the rank of the data matrix $\mbf{H}_k$. If $\mbf{H}_k$ has full row rank, then $l$ is increased till the matrix becomes row rank deficient. This is summarised in the following result.

\begin{proposition}
    Given a data matrix $\mbf{H}_k$ as in Eq.~\eqref{Eq:data_matrix}, the minimum number of time delayed states required to build the Hankel DMD/AR model is given by the rank of $\mbf{H}_k$, computed when $\mbf{H}_k$ is row rank deficient. Also,  $rank(\mbf{H}_k) = 2M $, where $M$ is the number of frequencies in the signal $\mbf{x}(k)$.
\end{proposition}
\textit{Proof:} Let $l^*$ be the minimum number of time delays required for the AR model. Then, one can predict the trajectory of the system in Eq.~\eqref{Eq:state_equation} using
\begin{align}
    \mbf{x}(k+l^*) & = \alpha_0 \mbf{x}(k) + \alpha_1 \mbf{x}(k+1) + \cdots +\alpha_{l^*-1} \mbf{x}(k+l^*-1). \label{Eq:AR_model_min}
\end{align}
Let the number of time delays $l$ used to build the data matrix $\mbf{H}_k$ be greater than the minimum number, i.e. $l > l^*$. Then, $\mbf{X}_{k+l^*}, \mbf{X}_{k+l^* +1}, \cdots, \mbf{X}_{k+l-1}$ will be a linear combination of $\mbf{X}_{k}, \mbf{X}_{k+1}, \cdots,  \mbf{X}_{k+l^*-1}$ due to Eq.~\eqref{Eq:AR_model_min}. Hence, $\mbf{H}_k$ will be row rank deficient, and its rank will give us the minimum number of time delays $l^*$. Owing to Proposition~\ref{prop:1}, $l^* = 2M$. \hfill $\blacksquare$ 
\begin{remark}
    In relation to the DMD algorithm, the above proposition translates to $rank(\mbf{H}_k) = rank(\mbf{A}) = rank(\bsym{\Tilde{A}}) = 2M $.
\end{remark}

\section{Results}
\label{Sec:Res}
The normalized 2-norm error ($\bsym{\epsilon}$) is used as the error metric to assess training and prediction accuracy. It is calculated between the true state data $\mbf{x}(k)$ and the estimates $\bsym{\hat{x}}(k)$ as given by, 
\begin{equation}
\bsym{\epsilon}({\mbf{x}}, \bsym{\hat{x}})=\frac{\|\bsym{\hat{x}}(k)-{\mbf{x}}(k)\|_2}{\max(\mbf{x})}
\end{equation}
It must be noted that $\bsym{\epsilon}$ is calculated separately for position and velocity states. The mean of the two error vectors is considered, given that the magnitude of position vectors is several orders of magnitude greater than velocity. This analysis involved the evaluation of the DMD algorithm using various state vector datasets generated for the two candidate satellites. The International Space Station (ISS) was chosen because of its popularity. A second object of interest, MOLNIYA-3-50, was considered due to its highly eccentric orbit, serving as a comparison against the ISS whose orbit is roughly circular. 
Training data is obtained by forward integrating the dynamical equations for the nonlinear pendulum, the Keplerian case, and non-Keplerian cases with perturbations. 
For SGP4, the analytic model is used with the mean orbital element set to generate propagations. The propagated mean elements are converted to osculating elements and corresponding state vectors before use. All orbital models are set in the Earth centered inertial (ECI) frame and use the True Equator Mean Equinox frame. To maintain consistency in training and testing across methods, fixed initial conditions based on the object's TLE are used. Data was generated at $0.01$ second intervals for the pendulum and at $1$ minute intervals for the orbital mechanics scenarios.
The evaluation criterion involved assessing the accuracy through normalized 2-norm errors across varying time delays, training window sizes ($W_{\text{TRN}}$), and their respective prediction windows ($W_{\text{PRED}}$). In the MOLNIYA-3-50 case, testing done on the SGP4 dataset alone is presented for conciseness.
Additionally, the spectral content determined in the signal by the DMD algorithm is evaluated against that obtained by the FFT. In this case, the FFT is used merely as an indicator to confirm that DMD has captured the spectral content of the signal in a gross manner. The experimental analysis was set up in the following manner. Initially, a constant training window size of $W_{\text{TRN}} = 10$ periods and a sampling interval ($\Delta k$) were selected for each case following the Nyquist rate. During this phase, the prediction region remained at $W_{\text{PRED}}=0$, ensuring DMD didn't extrapolate the signal. To determine the number of time delays, an exhaustive search involved multiple runs of the DMD algorithm, varying the time delays and selecting the run with the minimum error $\bsym{\epsilon}$. A similar search was conducted, this time adjusting the training window size while keeping the number of time delays constant as per the earlier step. Eventually, the prediction window size was determined by fixing the previous two model parameters at their established values. Selecting the sampling interval $\Delta k$ is largely experimental since the only known frequency is the orbit period. It requires a trial-and-error process to observe how alterations in the sampling interval affect both the chosen error metric and the frequency content of the reconstruction. More details on the sampling interval choice are discussed in Sec.~\ref{Sec:sampling}. The following sections discuss the results from each case.

\subsection{Nonlinear Oscillator}

\begin{figure}[h]
\begin{minipage}{.5\linewidth}
\raggedleft
\subfloat[]{\label{fig:TD_oscillator_main:a}\includegraphics[scale=.33]{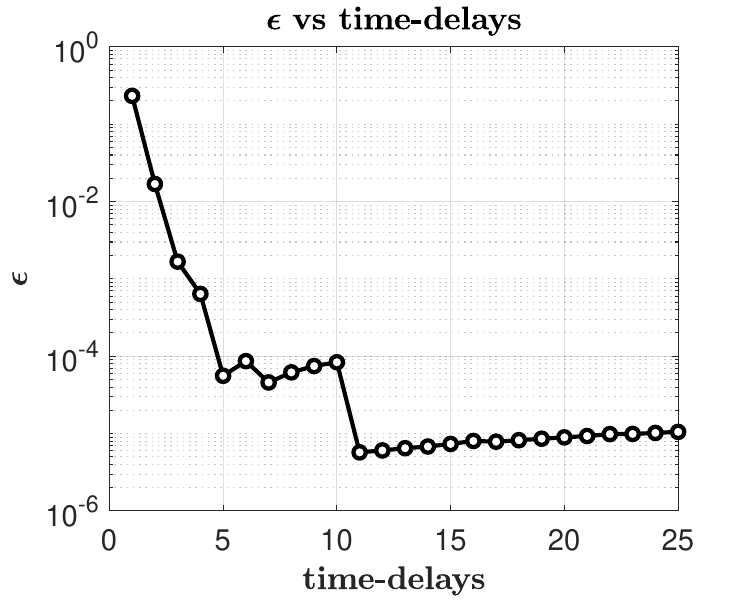}}
\end{minipage}%
\begin{minipage}{.5\linewidth}
\raggedright
\subfloat[]{\label{fig:TD_oscillator_main:d}\includegraphics[scale=.33]{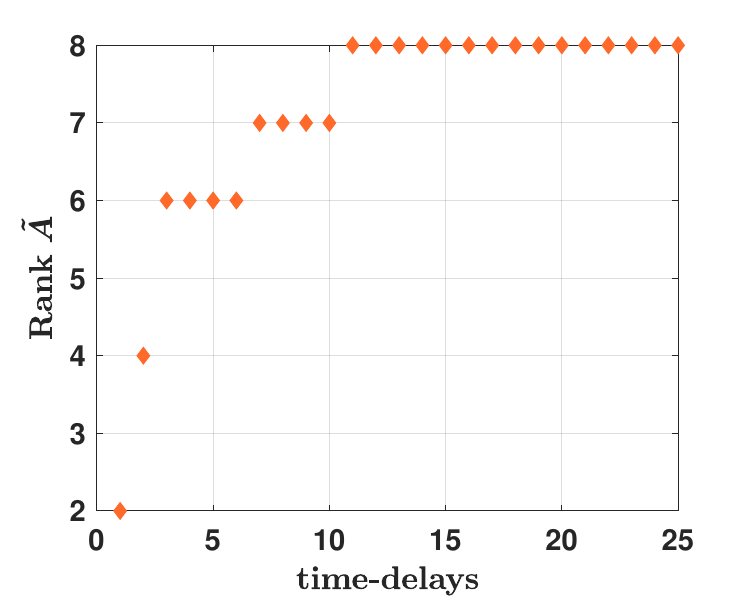}}
\end{minipage}
\begin{minipage}{.5\linewidth}
\raggedleft
\subfloat[]{\label{fig:TD_oscillator_main:b}\includegraphics[scale=.33]{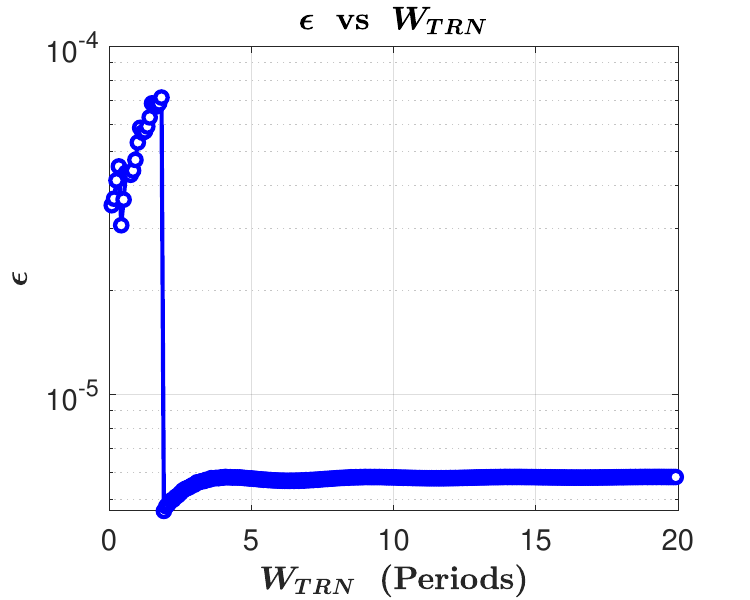}}
\end{minipage}
\begin{minipage}{.5\linewidth}
\raggedright
\subfloat[]{\label{fig:TD_oscillator_main:c}\includegraphics[scale=.33]{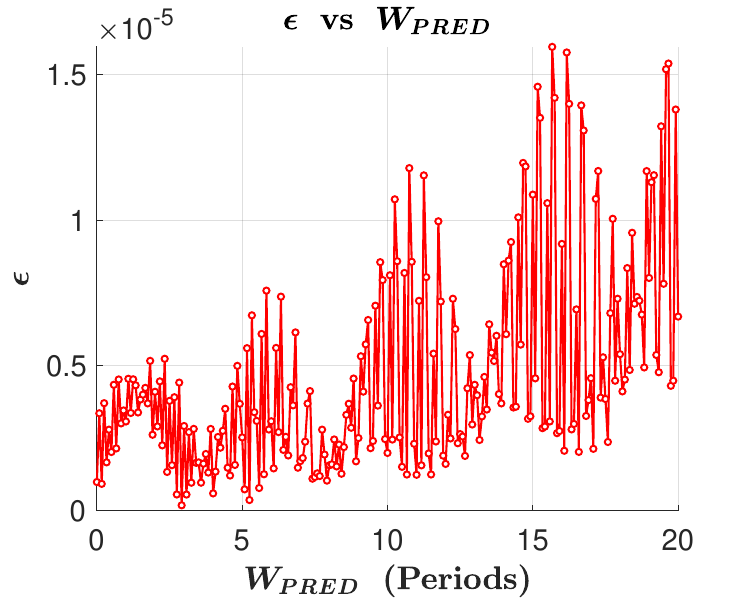}}
\end{minipage}
\caption{Nonlinear Oscillator model.}
\label{fig:TD_oscillator}
\end{figure}

The simple pendulum is considered as a toy example to illustrate the theory developed in Sec.~\ref{Sec:Th},
\begin{align}
    \ddot{\bsym{\theta}} = -\frac{g}{L}\sin\bsym{\theta}. \label{Eq:simple_pendulum}
\end{align}
For small angles, the motion can be expressed by an equivalent linear model. Here, the initial conditions are set to  $(\bsym{\theta}(0),\dot{\bsym{\theta}}(0)) = (\frac{\pi}{2}, 0)$, ensuring that nonlinearity prevails. Table.~\eqref{tab:oscillator} shows the frequencies identified by DMD and FFT. Here an arbitrarily chosen sampling interval of $\Delta k = 0.1$ seconds was used.  Based on the theory from Sec.~\ref{Sec:Th}, at least $8$ time delays are required to model the nonlinear oscillator. Fig.~\eqref{fig:TD_oscillator_main:a} shows that when $11$ time-delays are used the error $\bsym{\epsilon}$ is lowest. However, the rank of the reduced order matrix, $\bsym{\Tilde{A}}$ from Eq.~\eqref{eq:Atilde}, is 8. This is seen in Fig.~\eqref{fig:TD_oscillator_main:d} and it corresponds to $2$ times the number of frequencies identified by DMD. The discrepancy occurs because of the chosen sampling interval $\Delta k$. This is a common issue in real-world applications that employ a dataset of observations lacking guidance on the choice of $\Delta k$. Perfectly uncorrelated samples cannot be achieved in practice. Generally, it is recommended to use a rate greater than twice the Nyquist rate ~\cite{oppenheim1999discrete}. An alternative to address this issue is to start with time delays much larger than what is required and allow the DMD algorithm to identify dominant modes based on the SVD thresholding criterion ~\cite{brunton2022data}. 
Figs.~\eqref{fig:TD_oscillator_main:b} and ~\eqref{fig:TD_oscillator_main:c} show the respective sizes of $W_{\text{TRN}}$ and $W_{\text{PRED}}$. It must be noted the FFT does not show the higher frequencies. This observation is consistent across all test cases, as will be seen in the following sections, and is addressed in Sec.~\ref{Sec.FFT}.

\begin{table}[t]
\begin{tabular}{|c|c|c|c|}
\hline
$\mbf{\nu_{DMD}(mHz)}$ & $\mbf{\nu_{FFT}(mHz)}$ & $\mbf{\lambda_{DMD}}$   & $\mbf{|\lambda_{DMD}|}$ \\ \hline
0.8472                      & 0.8472                       & \begin{tabular}[c]{@{}c@{}}0.8616 + 0.5075i\\0.8616 - 0.5075i\end{tabular}   & \begin{tabular}[c]{@{}c@{}}1.0000\end{tabular} \\ \hline
2.5420                     &  2.5364                      & \begin{tabular}[c]{@{}c@{}}-0.0264 + 0.9996i\\ -0.0264 - 0.9996i\end{tabular} & \begin{tabular}[c]{@{}c@{}}1.0000\end{tabular} \\ \hline
4.0696                         &          -                   & \begin{tabular}[c]{@{}c@{}}-0.8337 + 0.5157i\\ -0.8337 - 0.5157i\end{tabular} & \begin{tabular}[c]{@{}c@{}}0.9998\end{tabular} \\ \hline
4.2361                          &          -                   & \begin{tabular}[c]{@{}c@{}}-0.8870 + 0.4617i\\ -0.8870 - 0.4617i\end{tabular} & \begin{tabular}[c]{@{}c@{}}1.0000\end{tabular} \\ \hline
\end{tabular}
\caption{Frequencies and Eigenvalues for the nonlinear oscillator model.}
\label{tab:oscillator}
\end{table}

\subsection{ISS Case Study}
\begin{table}[t]
\begin{tabular}{|c|c|c|c|c|c|}
\hline
$\mbf{a (km)}$ & $\mbf{e}$ & $\mbf{i(deg)}$ & $\mbf{\Omega(deg)}$ & $\mbf{\omega(deg)}$ & $\mbf{f(deg)}$ \\ \hline
6796.9         & 0.0007     & 51.639         & 113.73                            & 51.197                             & 358.89        \\ \hline
\end{tabular}
\caption{ISS initial orbital element set}
\label{tab:ISS_init}
\end{table}
This section discusses experimental observations and results from testing on the different datasets (with and without perturbations) generated for the ISS. 
Table~\eqref{tab:ISS_init} shows the initial conditions used, obtained from its TLE set. Appendix~\ref{App:A} shows the complete TLE set.




\subsubsection{Keplerian dynamics}

\begin{align}
\ddot{\mbf{r}}=-\mu_{\oplus} \frac{\mbf{r}}{r^3}+\mbf{p}
\label{Eq:Kep_no_p}
\end{align}

\begin{figure}[h]
\begin{minipage}{.5\linewidth}
\raggedleft
\subfloat[]{\label{fig:TD_Kep:main:a}\includegraphics[scale=.42]{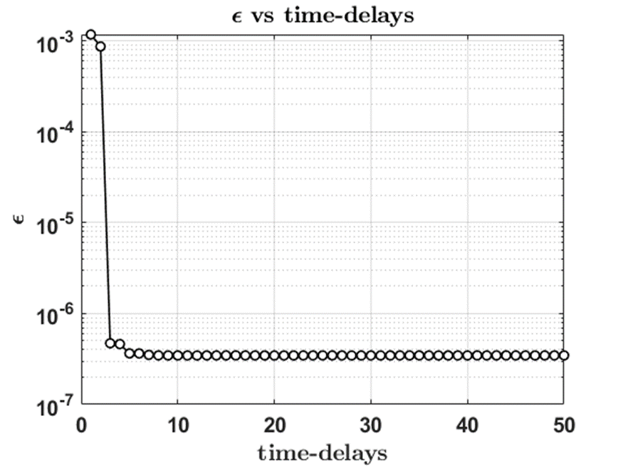}}
\end{minipage}%
\begin{minipage}{.5\linewidth}
\raggedright
\subfloat[]{\label{fig:TD_Kep:main:b}\includegraphics[scale=.42]{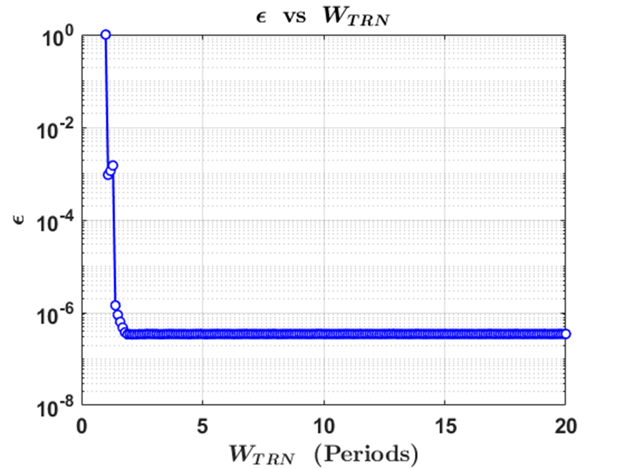}}
\end{minipage}
\begin{minipage}{.5\linewidth}
\raggedleft
\subfloat[]{\label{fig:TD_Kep:main:d}\includegraphics[scale=.42]{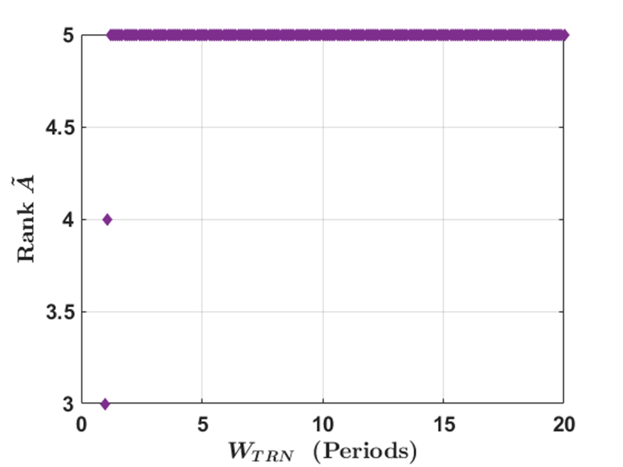}}
\end{minipage}
\begin{minipage}{.5\linewidth}
\raggedright
\subfloat[]{\label{fig:TD_Kep:main:c}\includegraphics[scale=.42]{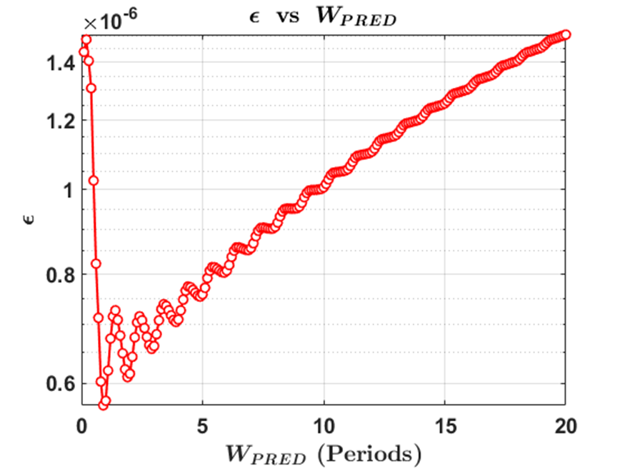}}
\end{minipage}
\caption{ISS (Keplerian dynamics)}
\label{fig:TD_Kep}
\end{figure}

In non-Keplerian mechanics, the motion of a space object is described in three degrees of freedom (3DOF) by the two-body problem subject to disturbing forces, as seen in equation Eq.~\eqref{Eq:Kep_no_p}. Here $\mbf{p}$ represents various disturbance accelerations arising from the central body, the atmosphere, interactions with other bodies, solar radiation pressure, and other perturbations. In Eq.~\eqref{Eq:Kep_no_p}, $\mbf{r}$ is the relative position vector of the satellite with respect to Earth, assuming point masses, and $\mu_{\oplus}$ is Earth's gravitational parameter. In this subsection, we first consider two-body Keplerian dynamics without any perturbations, i.e., $\mbf{p} = \mbf{0}$. A sampling interval of $\Delta k = 11$ minutes was used. Table~\eqref{tab:ISS_Kep} shows the frequencies identified by the DMD algorithm and FFT algorithms respectively with their corresponding eigenvalues and magnitudes. It must be noted the FFT does not show the higher harmonics, this issue is addressed in Sec.~\ref{Sec.FFT}. Fig.~\eqref{fig:TD_Kep} shows the results obtained for the choice in the number of time delays, training window length, and maximum prediction window. From DMD analysis, it is seen that there is a fundamental frequency representing the period of the object, a higher harmonic which is $2$ times the fundamental value, and a real mode exactly on the unit circle. In this case, based on the theory developed in Sec.~\ref{Sec:Th}, a total of $5$ time delays are required, two for each frequency and one for the real mode. This is verified by the results in Fig.~\eqref{fig:TD_Kep:main:a}.  Fig.~\eqref{fig:TD_Kep:main:b} shows that a small training window of about $1.3$ periods is sufficient. However, it's essential to ensure a sufficiently long training window such that the $rank(\bsym{\Tilde{A}}) = 5$, where $\bsym{\Tilde{A}}$ is the reduced order matrix from Eq.~\eqref{eq:Atilde}. Fig.~\eqref{fig:TD_Kep:main:d} show the change in $rank(\bsym{\Tilde{A}})$ as the size of the training window is increased. A further increase in size, beyond $1.3$ periods does not change the rank of $\bsym{\Tilde{A}}$ but does improve the conditioning of the input matrix $\mbf{X_k}$ thus improving $\bsym{\epsilon}$. Fig.~\eqref{fig:TD_Kep:main:c} shows roughly the same error magnitude in $W_{\text{PRED}}$ for $W_{\text{TRN}} = 10$. This is because Keplerian dynamics are perfectly periodic and unforced, allowing DMD to capture it well.

\begin{table}[t]
\begin{tabular}{|c|c|c|c|}
\hline
$\mbf{\nu_{DMD}(mHz)}$ & $\mbf{\nu_{FFT}(mHz)}$ & $\mbf{\lambda_{DMD}}$   & $\mbf{|\lambda_{DMD}|}$ \\ \hline
0.1793                       & 0.1793                       & \begin{tabular}[c]{@{}c@{}}0.7360 + 0.6769i\\ 0.7360 - 0.6769i\end{tabular}   & \begin{tabular}[c]{@{}c@{}}1.0000\end{tabular} \\ \hline
0.3586                       & -                       & \begin{tabular}[c]{@{}c@{}}0.0834 + 0.9965i\\ 0.0834 - 0.9965i\end{tabular} & \begin{tabular}[c]{@{}c@{}}1.0000\end{tabular} \\ 
\hline
-                       & -                       & \begin{tabular}[c]{@{}c@{}}1.000 + 0.0000i\\ \end{tabular} & \begin{tabular}[c]{@{}c@{}}1.0000\end{tabular} \\ \hline
\end{tabular}
\caption{Frequencies and Eigenvalues for ISS modeled by Keplerian dynamics}
\label{tab:ISS_Kep}
\end{table}

\subsubsection{Non-Keplerian dynamics with J2}
\begin{figure}[h]
\begin{minipage}{.5\linewidth}
\raggedleft
\subfloat[]{\label{fig:TD_Kep_J2:main:a}\includegraphics[scale=.42]{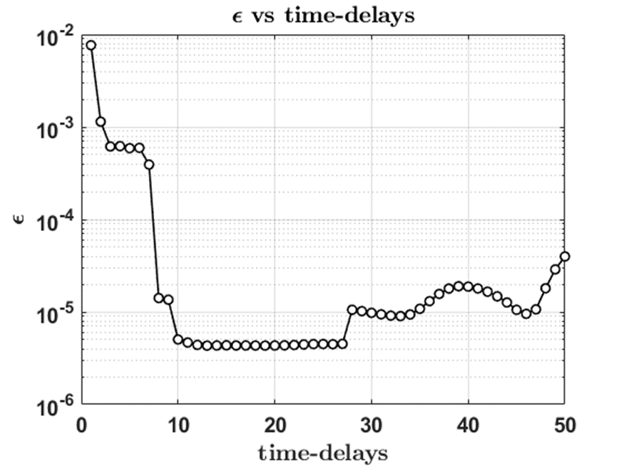}}
\end{minipage}%
\begin{minipage}{.5\linewidth}
\raggedright
\subfloat[]{\label{fig:TD_Kep_J2:main:b}\includegraphics[scale=.42]{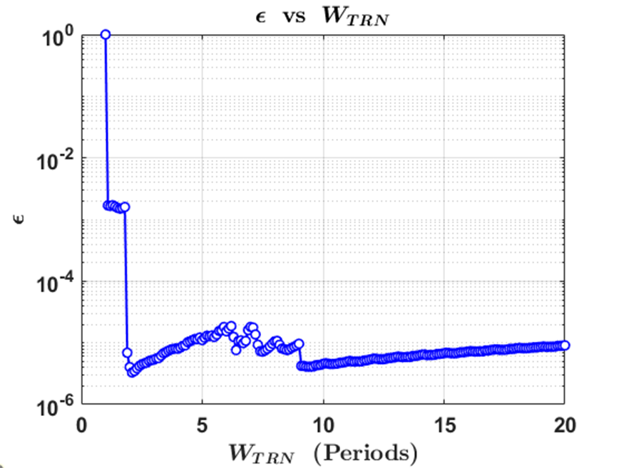}}
\end{minipage}
\begin{minipage}{.5\linewidth}
\centering
\subfloat[]{\label{fig:TD_Kep_J2:main:d}\includegraphics[scale=.42]{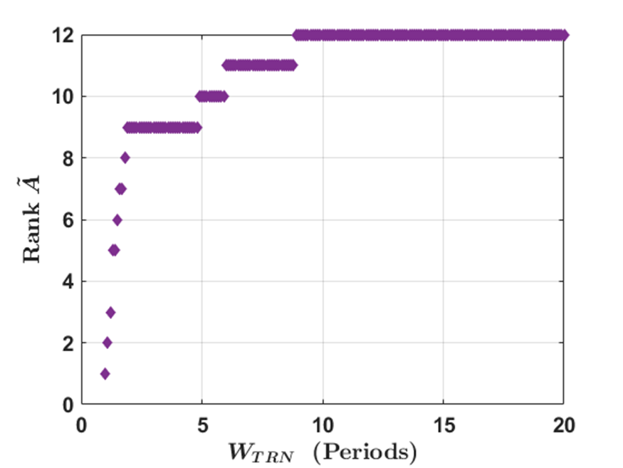}}
\end{minipage}
\begin{minipage}{.5\linewidth}
\raggedright
\subfloat[]{\label{fig:TD_Kep_J2:main:c}\includegraphics[scale=.42]
{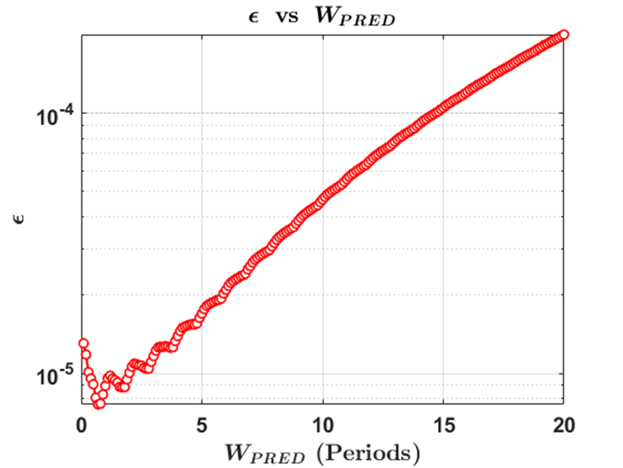}}
\end{minipage}
\caption{ISS (Non-Keplerian Dynamics with J2)}
\label{fig:TD_Kep_J2}
\end{figure}

Consider a dynamic model at ``sufficiently high'' altitude, i.e., with negligible impact of the atmosphere, but also sufficiently low, such that the dominant perturbative acceleration is due to Earth's oblateness, captured in the form of a  modified potential field,
\begin{gather}
\ddot{\mbf{r}}=-\mu_{\oplus} \frac{\mbf{r}}{r^3}+\mbf{p_{J2}} \ \text{,where,} \label{Eq:Kep_J2}\\
 \mbf{p_{J2}} = \frac{3 J_2 \mu_{\oplus} R^2}{2}\left[\frac{x}{r}\left(5 \frac{z^2}{r^2}-1\right) \hat{\mbf{i}}+\frac{y}{r}\left(5 \frac{z^2}{r^2}-1\right) \hat{\mbf{j}}+\frac{z}{r}\left(5 \frac{z^2}{r^2}-3\right) \hat{\mbf{k}}\right] \nonumber
\end{gather}
Among the zonal harmonics, the first harmonic J2 significantly dominates this effect by approximately three orders of magnitude compared to higher-degree harmonics. J2 perturbations cause gradual changes in three orbital elements: the Right Ascension of the Ascending Node (RAAN) ($\Omega$), the argument of perigee ($\omega$), and the mean anomaly. Remarkably, the semi-major axis, eccentricity, and inclination experience minimal changes on average over a full period due to Earth's zonal oblateness. It's important to note that these are averaged effects, implying that perturbations above the mean are balanced by perturbations below the mean within the span of a period. Eq.~\eqref{Eq:Kep_J2} describes the dynamics of space objects subject to J2 perturbations, where $\mbf{p_{J2}}$ is the perturbing acceleration in the cartesian coordinate system expressed in terms of unit vectors $\bsym{(\hat{i}, \hat{j}, \hat{k})}$.  In this case, the dataset was sampled at $\Delta k = 7$ minutes. DMD has identified $6$ frequencies, as seen in table~\eqref{tab:ISS_Kep_J2}, implying that the number of time delays required is $12$ which agrees well with Fig.~\eqref{fig:TD_Kep_J2:main:a}. Figs.~\eqref{fig:TD_Kep_J2:main:b} and ~\eqref{fig:TD_Kep_J2:main:c} show the corresponding sizes for training and prediction windows respectively, $W_{\text{TRN}}$ and $W_{\text{PRED}}$. Fig.~\eqref{fig:TD_Kep_J2:main:d} shows the variation in $rank(\bsym{\Tilde{A}})$ with increasing training window size. Notably, when $\bsym{\Tilde{A}}$ has not reached the full required rank of $12$, the error values are still quite low. This discrepancy is mainly due to the significantly smaller modal amplitude corresponding to $\nu_{\text{DMD}} = 1.043 \times 10^{-7}$ mHz compared to other frequencies. Modal amplitude is denoted by $\mbf{b}$ in Eq.~\eqref{eq:DMD_time_dynamics}. In this instance, $W_{\text{TRN}} \geq 8.9$ periods prove to be sufficient. Fig.~\eqref{fig:TD_Kep_J2:main:c} shows error $\mbf{\epsilon}$ in $W_{\text{PRED}}$ for $W_{\text{TRN}} = 8.9$.
Furthermore, it can be seen that predictions become less accurate as the size of the extrapolation region increases. This is consistent with the understanding that DMD is approximating a forced nonlinear system with a higher dimensional linear-time invariant system. The frequency content in the signal is continuously changing and the discrete spectra identified by DMD cannot capture this change.

\begin{table}[t]
\begin{tabular}{|c|c|c|c|}
\hline
$\mbf{\nu_{DMD}(mHz)}$ & $\mbf{\nu_{FFT}(mHz)}$ & $\mbf{\lambda_{DMD}}$   & $\mbf{|\lambda_{DMD}|}$ \\ \hline
1.043e-07                       & -                      & \begin{tabular}[c]{@{}c@{}}0.9999 + 2.7522e-04i\\ 0.9999 - 2.7522e-04i\end{tabular}   & \begin{tabular}[c]{@{}c@{}}0.9999\end{tabular} \\ \hline
0.1793                       & 0.1793                       & \begin{tabular}[c]{@{}c@{}}0.8901 + 0.4556i\\ 0.8901 - 0.4556i\end{tabular}   & \begin{tabular}[c]{@{}c@{}}0.9999\end{tabular} \\ \hline
0.1794                       & -                       & \begin{tabular}[c]{@{}c@{}}0.8899 + 0.4559i\\ 0.8899 - 0.4559i\end{tabular}   & \begin{tabular}[c]{@{}c@{}}0.9999\end{tabular} \\ \hline
0.1797                       & -                       & \begin{tabular}[c]{@{}c@{}}0.8897 + 0.4564i\\ 0.8897 - 0.4564i\end{tabular}   & \begin{tabular}[c]{@{}c@{}}1.0000\end{tabular} \\ \hline
0.3586                       & -                       & \begin{tabular}[c]{@{}c@{}}0.5846 + 0.8112i\\ 0.5846 - 0.8112i\end{tabular} & \begin{tabular}[c]{@{}c@{}}1.0000\end{tabular} \\ \hline
0.5381                       & -                       & 
\begin{tabular}[c]{@{}c@{}}0.1501 + 0.9887i\\ 0.1501 - 0.9887i\end{tabular}   & \begin{tabular}[c]{@{}c@{}}1.0000\end{tabular} \\ \hline                                      
\end{tabular}
\caption{Frequencies and Eigenvalues for ISS modeled by non-Keplerian dynamics with J2 perturbation}
\label{tab:ISS_Kep_J2}
\end{table}

\subsubsection{Non-Keplerian Dynamics with Drag}
\begin{figure}[h]
\begin{minipage}{.5\linewidth}
\raggedleft
\subfloat[]{\label{fig:TD_Kep_D_main:a}\includegraphics[scale=.42]{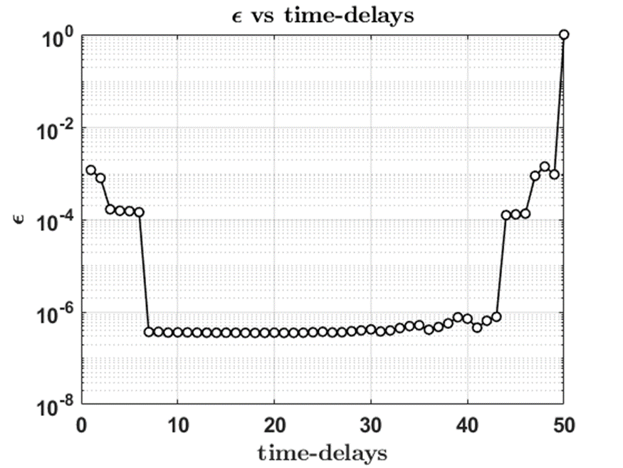}}
\end{minipage}%
\begin{minipage}{.5\linewidth}
\raggedright
\subfloat[]{\label{fig:TD_Kep_D_main:b}\includegraphics[scale=.42]{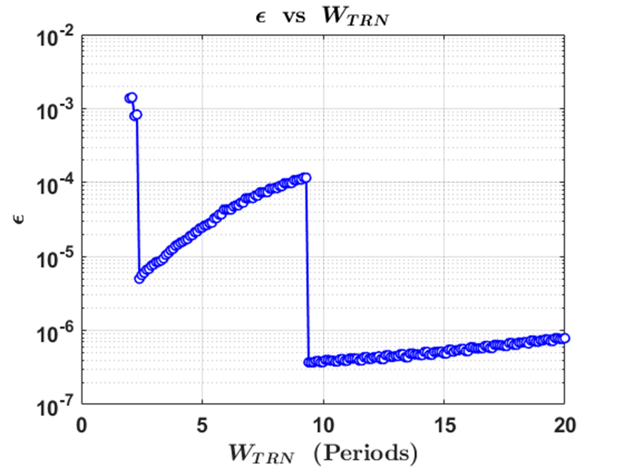}}
\end{minipage}
\begin{minipage}{.5\linewidth}
\raggedleft
\subfloat[]{\label{fig:TD_Kep_D_main:d}\includegraphics[scale=.42]{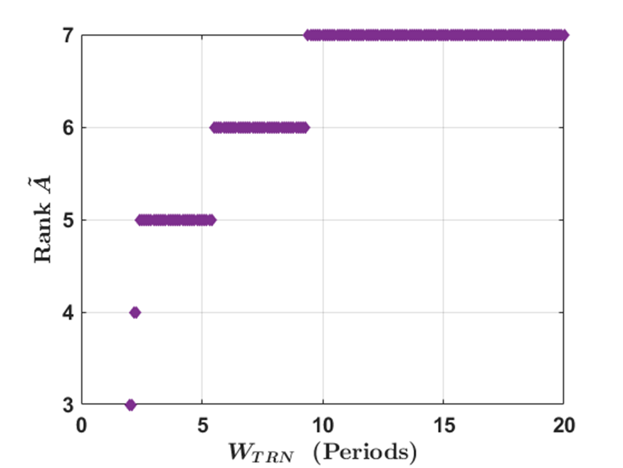}}
\end{minipage}
\begin{minipage}{.5\linewidth}
\raggedright
\subfloat[]{\label{fig:TD_Kep_D_main:c}\includegraphics[scale=.42]{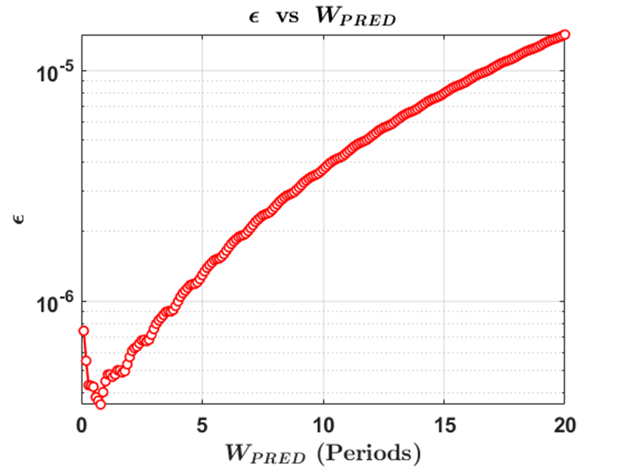}}
\end{minipage}
\caption{ISS (Non-Keplerian dynamics with Drag)}
\label{fig:TD_Kep_D}
\end{figure}

In this section, we consider a dynamic model where the dominant 
perturbation is drag, which causes a gradual decay of the orbit. A simplified drag model is used assuming continuous atmosphere as follows,
\begin{gather}
\ddot{\mbf{r}}=-\mu_{\oplus} \frac{\mbf{r}}{r^3}+\mbf{p_D} \ 
\text{, where,}  \label{Eq:Kep_D} \\
\mbf{p_D} =-\frac{1}{2} \rho v_{\mathrm{rel}}\left(\frac{C_D A}{m}\right) \mbf{v}_{\mathrm{rel}} \nonumber
\end{gather}
In terms of the ballistic parameter, $B^*$, obtained from the TLE, $\mbf{p_D} = -\frac{\rho}{\rho_0} B^{*}v_{rel}\mbf{v}_{\mathrm{rel}}$, where $B^{*} = \frac{\rho_0 C_d A}{2m}$.
The extent of drag's impact varies based on various factors, including the orientation of the satellite and the conditions encountered within the atmosphere (such as diurnal, solar, and geomagnetic variations). Drag induces secular changes in the semi-major axis ($a$), eccentricity ($e$), and inclination ($i$) orbital elements. 
Eq.~\eqref{Eq:Kep_D} describes the motion of space objects subject to Drag. In this case, $\Delta k = 18$ minutes was used to sample the dataset.
Table~\eqref{tab:ISS_Kep_Drag} shows the frequencies obtained by the DMD algorithm. Fig.~\eqref{fig:TD_Kep_D} shows the results obtained from the experimental analysis, it is consistent with the observations from the previous cases. A total of $7$ time delays are required to account for the $3$ frequencies and $1$ real mode found. Figs.~\eqref{fig:TD_Kep_D_main:b} and ~\eqref{fig:TD_Kep_D_main:c} show the sizes for $W_{TRN}$ and $W_{\text{PRED}}$ respectively. Here, $W_{\text{TRN}} \geq 9.4$ periods is required here to ensure that ${\bsym{\Tilde{A}}}$, the reduced order matrix from Eq.~\eqref{eq:Atilde}, reaches the requisite rank. 
This is demonstrated in Fig. ~\eqref{fig:TD_Kep_D_main:d}. Fig.~\eqref{fig:TD_Kep_D_main:c} shows error $\bsym{\epsilon}$ in $W_{\text{PRED}}$ for $W_{\text{TRN}} = 9.4$ periods.

\begin{table}[t]
\begin{tabular}{|c|c|c|c|}
\hline
$\mbf{\nu_{DMD}(mHz)}$ & $\mbf{\nu_{FFT}(mHz)}$ & $\mbf{\lambda_{DMD}}$   & $\mbf{|\lambda_{DMD}|}$ \\ \hline
0.1793                       & 0.1793                       & \begin{tabular}[c]{@{}c@{}}0.3475 + 0.9383i\\ 0.3475 - 0.9383i\end{tabular}   & \begin{tabular}[c]{@{}c@{}}1.0006\end{tabular} \\ \hline
0.1794                       & -                       & \begin{tabular}[c]{@{}c@{}}0.3458 + 0.9375i\\ 0.3458 - 0.9375i\end{tabular}   & \begin{tabular}[c]{@{}c@{}}0.9929\end{tabular} \\ \hline
0.3586                       & -                       & \begin{tabular}[c]{@{}c@{}}-0.7596 + 0.6503i\\ -0.7596 - 0.6503i\end{tabular} & \begin{tabular}[c]{@{}c@{}}1.0000\end{tabular} \\ \hline
-                            & -                            & 1.0000 + 0.0000i                                                              & 1.0000                                                  \\ \hline
\end{tabular}
\caption{Frequencies and Eigenvalues for ISS Modeled by Non-Keplerian Dynamics with Drag Perturbation}
\label{tab:ISS_Kep_Drag}
\end{table}

\subsubsection{Simplified General Pertubations model (SGP4)}
\label{sec:ISS_SGP4}
\begin{figure}[h]
\begin{minipage}{.5\linewidth}
\raggedleft
\subfloat[]{\label{fig:TD_SGP4:main:a}\includegraphics[scale=.42]{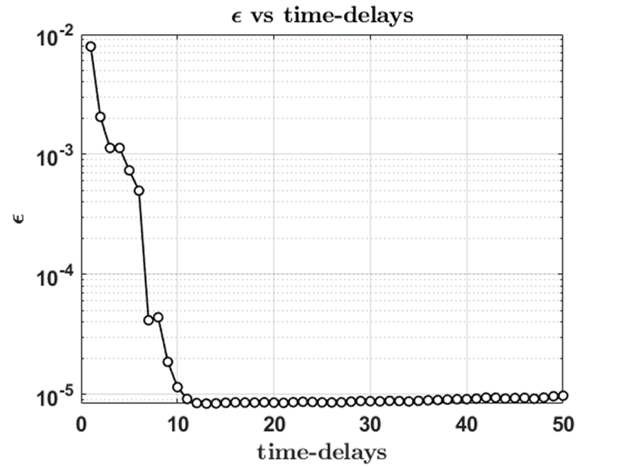}}
\end{minipage}%
\begin{minipage}{.5\linewidth}
\raggedright
\subfloat[]{\label{fig:TD_SGP4:main:b}\includegraphics[scale=.42]{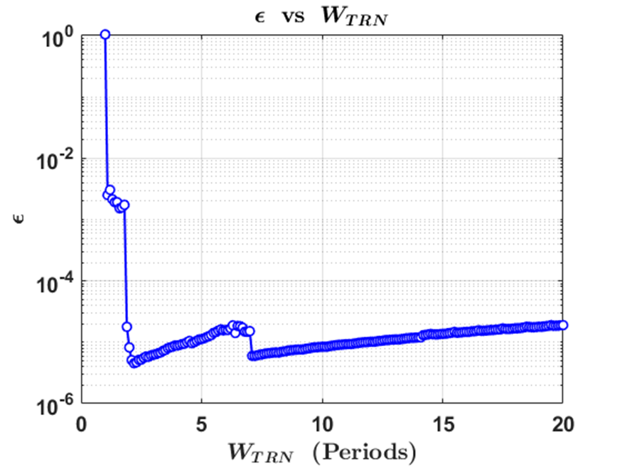}}
\end{minipage}
\begin{minipage}{.5\linewidth}
\raggedleft
\subfloat[]{\label{fig:TD_SGP4:main:d}\includegraphics[scale=.42]{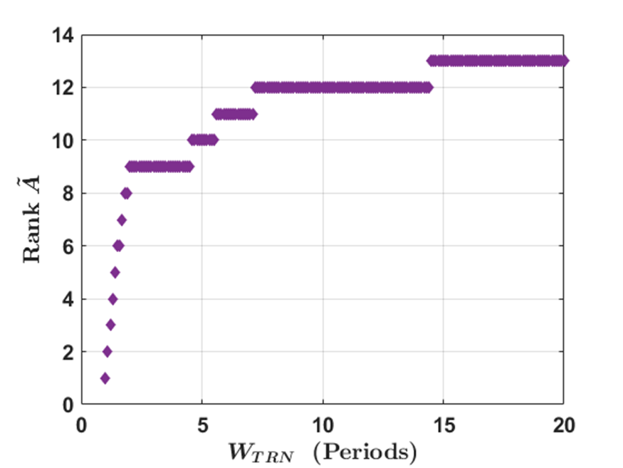}}
\end{minipage}
\begin{minipage}{.5\linewidth}
\centering
\subfloat[]{\label{fig:TD_SGP4:main:c}\includegraphics[scale=.42]{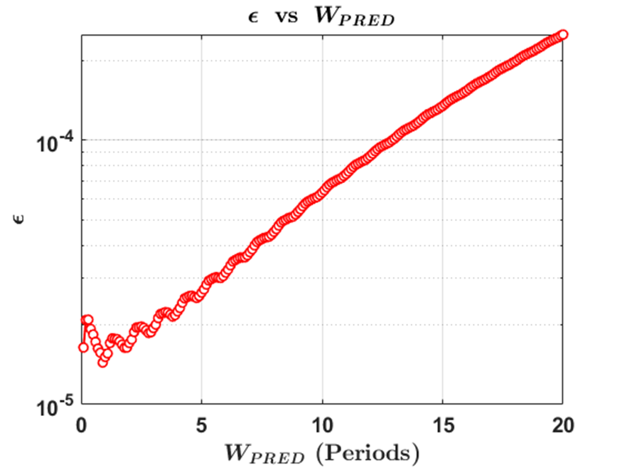}}
\end{minipage}
\caption{ISS (SGP4)}
\label{fig:TD_SGP4}
\end{figure}

Simplified analytic perturbation models are commonly employed to describe non-Keplerian dynamics dynamics in Earth-orbiting systems. Among these SGP4/SDP4 is recommended for use when propagating TLE sets. Computations are done within the ECI coordinate system, using the True Equator Mean Equinox reference frame ~\cite{hoots1980models}. The models consider secular and periodic variations due to Earth's oblateness, solar and lunar gravitational effects, gravitational resonance effects, and orbital decay using an atmospheric drag model ~\cite{cranford1969improved}. The above models perform  propagation in the mean orbital elements formulation, however, a conversion is made to osculating elements and respective state vectors. Here SGP4/SDP4 models are used for generating baseline trajectories, from which training datasets are built for learning satellite orbits. In this case, a sampling interval of $\Delta k = 7$ minutes was used. DMD identified $6$ frequencies, as seen in Table.~\eqref{tab:ISS_SGP4}. Fig.~\eqref{fig:TD_SGP4:main:a} shows that a total of $12$ time delays are sufficient. 
Figs.~\eqref{fig:TD_SGP4:main:b} and ~\eqref{fig:TD_SGP4:main:c} show the size of $W_{\text{TRN}}$ and $W_{\text{PRED}}$ which agree with observations thus far. Fig.~\eqref{fig:TD_SGP4:main:d} shows the change in $rank(\bsym{\Tilde{A}})$ as more training data is considered, here $14.1 \geq W_{\text{TRN}} \geq 7.2$ periods are sufficient. Fig.~\eqref{fig:TD_SGP4:main:c} shows error $\bsym{\epsilon}$ in $W_{\text{PRED}}$ for $W_{\text{TRN}} = 10$ periods. However, when the experiment was set up, the initial assumption considered only $10$ periods, aiming to determine the necessary number of time delays within that window. As the size of this training region increases beyond $10$ periods, an additional frequency becomes apparent, observed through the rise in $rank(\bsym{\Tilde{A}})$ to 13. The complexity of SGP4 models contributes to this change in frequency spectrum due to the continuous variation in forcing terms acting on the object. In such scenarios, an assumption can be made such that the initial guess for the number of time delays and the training window size is much larger than what is required. This approach allows the DMD algorithm to appropriately identify the suitable size for the reduced-order matrix $\bsym{\Tilde{A}}$ using the SVD thresholding criterion.

\begin{table}[t]
\begin{tabular}{|c|c|c|c|}
\hline
$\mbf{\nu_{DMD}(mHz)}$ & $\mbf{\nu_{FFT}(mHz)}$ & $\mbf{\lambda_{DMD}}$   & $\mbf{|\lambda_{DMD}|}$ \\ \hline
0.7119e-07 & - & \begin{tabular}[c]{@{}c@{}}0.9992 + 1.8787e-04i\\ 0.9992 - 1.8787e-04i\end{tabular} & 0.9999   \\ \hline
0.1792 & 0.1792 & \begin{tabular}[c]{@{}c@{}}0.8909 + 0.4556i\\ 0.8909 - 0.4556i\end{tabular}   & 1.0007   \\ \hline
0.1794 & -      & \begin{tabular}[c]{@{}c@{}}0.8891 + 0.4561i\\ 0.8891 - 0.4561i\end{tabular}   & 0.9999   \\ \hline
0.1798 & -      & \begin{tabular}[c]{@{}c@{}}0.8886 + 0.4565i\\ 0.8886 - 0.4565i\end{tabular}   & 0.9992   \\ \hline
0.3587 & -   & \begin{tabular}[c]{@{}c@{}}0.5844 + 0.8114i\\ 0.5844 - 0.8114i\end{tabular}   & 1.0001   \\ \hline
0.5381 & - & \begin{tabular}[c]{@{}c@{}}0.1500 + 0.9886i\\ 0.1500 - 0.9886i\end{tabular} & 0.9999   \\ \hline
\end{tabular}
\caption{Frequencies and Eigenvalues for ISS modeled by SGP4}
\label{tab:ISS_SGP4}
\end{table}

\subsection{Note on FFT comparison}
\label{Sec.FFT}
\begin{figure}
    \centering
    \includegraphics[scale=0.30]{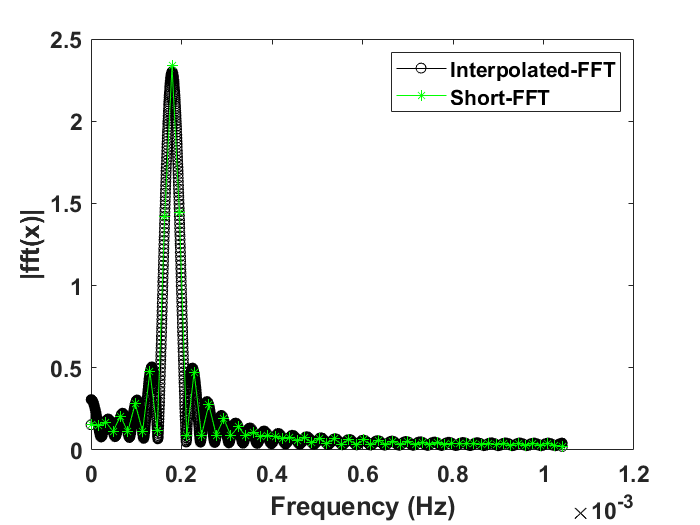}
    \caption{Comparison between interpolated FFT and short-FFT for the ISS SGP4 scenario}
    \label{fig:FFT_comp}
\end{figure}
The fast Fourier transform (FFT) is used only as an indicator to verify if the DMD algorithm captures the spectral content of the dataset crudely.
The FFT encounters challenges due to the small size of the dataset. The small size is in turn due to  the chosen sampling interval over the training window ($W_{\text{TRN}}$=10 periods). This limitation makes it challenging to accurately estimate higher harmonics solely through visual inspection. Fig.~\eqref{fig:FFT_comp} shows a comparison of the FFT plots for the "short" FFT performed on the input dataset and the interpolated FFT generated by zero-padding the signal. The ISS SGP4 dataset was used to generate the plots. The side lobes are artefacts that can be attributed to the basis vectors of the DFT not being exactly integer periodic within the bin width. This is primarily the effect of rectangular windowing where the leftover energy goes into the side lobes and is referred to as spectral leakage ~\cite{lyon2009discrete}.
A possible mitigation would be to use long windows of data spanning several days of orbital data and include frequent observations, but this is unrealistic in the present context. Windows such as the Hamming or Kaiser windows may be used to suppress the side lobes but they will not necessarily help with unraveling the higher harmonics. This also complicates the selection of $\Delta k$. Ideally, one would take the FFT of the signal within the window under consideration ($W_{\text{TRN}}$) and determine $\Delta k$ based on the highest frequency to be captured. However, this is not the case as only the period of the orbit is seen and the higher harmonics are not visible in the FFT.

\subsection{Molniya-3-50: The Case of a Highly Eccentric Orbit}
\begin{table}[t]
\begin{tabular}{|c|c|c|c|c|c|}
\hline
$\mbf{a (km)}$ & $\mbf{e}$ & $\mbf{i(deg)}$ & $\mbf{\Omega(deg)}$ & $\mbf{\omega(deg)}$ & $\mbf{f(deg)}$ \\ \hline
26555.94         & 0.7294     & 63.324         & 295.46                            & 282.69                             & 357.32        \\ \hline
\end{tabular}
\caption{MOLNIYA-3-50 Initial Orbital Element Set}
\label{tab:ML350_init}
\end{table}

\begin{figure}[h]
\begin{minipage}{.5\linewidth}
\raggedleft
\subfloat[]{\label{fig:TD_SGP4_ML350:main:a}\includegraphics[scale=.42]{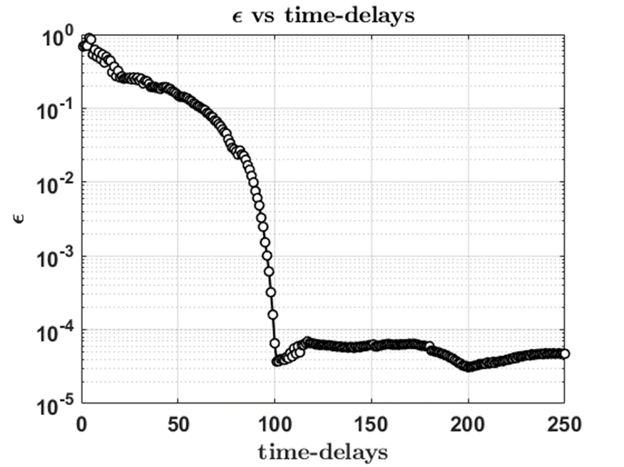}}
\end{minipage}%
\begin{minipage}{.5\linewidth}
\raggedright
\subfloat[]{\label{fig:TD_SGP4_ML350:main:b}\includegraphics[scale=.42]{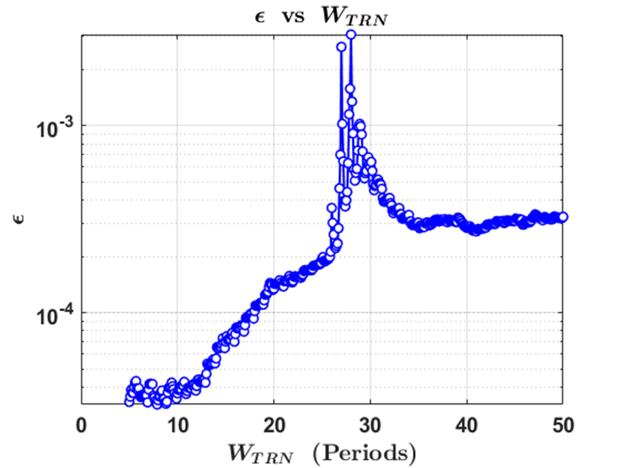}}
\end{minipage}
\begin{minipage}{.5\linewidth}
\raggedright
\subfloat[]{\label{fig:TD_SGP4_ML350:main:d}\includegraphics[scale=.42]{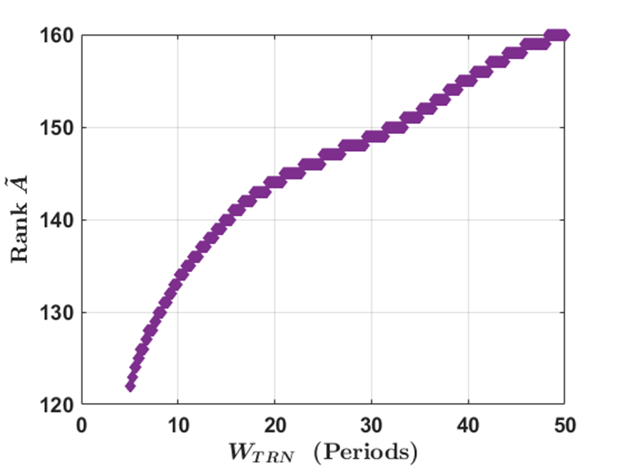}}
\end{minipage}
\begin{minipage}{.5\linewidth}
\raggedright
\subfloat[]{\label{fig:TD_SGP4_ML350:main:c}\includegraphics[scale=.42]{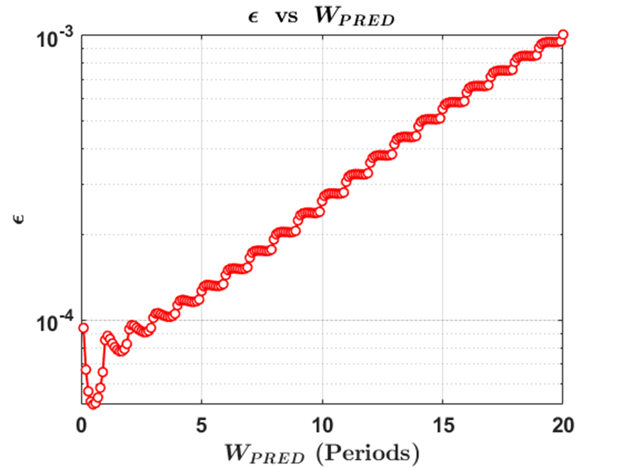}}
\end{minipage}
\caption{MOLNIYA-3-50 (SGP4)}
\label{fig:TD_SGP4_ML350}
\end{figure}

The above analysis was also performed on the Molniya-3-50 satellite, including all orbital perturbations modeled by SGP4/SDP4. The process was similar to that performed for the ISS. Following Sec.~\ref{Sec:sampling}, a constant training window size of $W_{\text{TRN}} = 10$ periods and a sampling interval of $\Delta k = 7$ minutes were used. The prediction region was set to $W_{\text{PRED}}=0$ while determining the number of time delays and training window size.  Finally, the size for the prediction window was found by setting the previous two model parameters to the constant values found. Fig.~\eqref{fig:TD_SGP4:main:a} shows that the number of time delays required exceeds 100, showing that the spectral content in this case is dense. Figs.~\eqref{fig:TD_SGP4:main:b} and ~\eqref{fig:TD_SGP4_ML350:main:c} show the respective sizes for $W_{\text{TRN}}$ and $W_{\text{PRED}}$. The initial assumption of using only $10$ periods to identify the number of time delays fails here. As Fig.~\eqref{fig:TD_SGP4_ML350:main:a} shows that a value of 100 should suffice, however, in Fig.~\eqref{fig:TD_SGP4_ML350:main:d} we see the growing rank of $\bsym{\Tilde{A}}$ as the training region size is gradually increased. This discrepancy is because the frequency content of the dataset is constantly changing. A similar observation was made in Sec.~\ref{sec:ISS_SGP4}, where the increase in training window size beyond $10$ periods introduced new frequencies in the spectrum. Here, it can be seen that error in the $W_{\text{TRN}}$ does saturate, but, the rank of $\bsym{\Tilde{A}}$ is still growing. 
Fig.~\eqref{fig:TD_SGP4_ML350:main:c} shows $\mbf{\epsilon}$ error in $W_{\text{PRED}}$ for $W_{\text{TRN}} = 10$ periods and 101 time delays. 
It is worth mentioning that, the magnitude of $\bsym{\epsilon}$ is higher than the values observed for the ISS case. 
This difference in behavior can be attributed to the orbits' eccentricity and the time scales involved. In the case of the ISS, $10$ periods correspond to approximately $1.5$ hours, whereas for MOLNIYA-3-50, it equates to roughly $5$ days. This discrepancy allows ample time for perturbations to significantly impact the trajectory. The perturbations cause drift in the dynamics, thus making it difficult for DMD to approximate the frequency spectrum. The comparison between ISS and MOLNIYA-3-50 is drawn to illustrate the substantial difference in applications and how the use of DMD demands a meticulous, case-specific approach rather than a generalized one.

\subsection{Effect of Sampling Rate}
\label{Sec:sampling}
\begin{figure}
    \centering
    \includegraphics[scale=0.35]{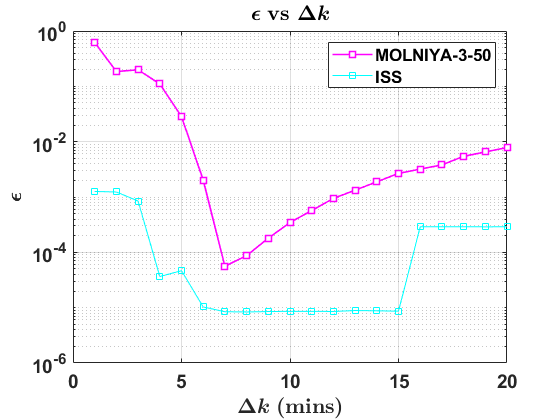}
    \caption{Effect of different sampling intervals}
    \label{samp_effect_orbital}
\end{figure}
From the experiments, it was observed that DMD is very sensitive to the sampling interval between input time snapshots.  In general, it is sufficient to use a sampling rate at least twice the highest frequency within the signal, following the lower bound specified by the Nyquist criterion ~\cite{schmid2010dynamic}. However, finding the Nyquist frequency is challenging. A potential approach is to start with a considerably large number of time delays and let the DMD algorithm identify the necessary number of DMD modes for reconstruction.
The number of DMD modes is given by the $rank(\bsym{\Tilde{A}})$ or $r$ identified by the SVD thresholding algorithm.
If the number of identified DMD modes is less than the initial number of time delays, a gradual increase in the sampling interval can be applied to test for sensitivity and pinpoint the $\Delta k$ that yields the minimum $\bsym{\epsilon}$. However, if the number of DMD modes equals the initial number of time delays, this process needs to be repeated by increasing the number of time delays until the DMD modes reach a saturation point. The sensitivity test to identify $\Delta k$ such that $\bsym{\epsilon}$ is minimum can be performed then. An example of this test is as follows. Fig.~\eqref{samp_effect_orbital} shows a comparison of the impact of changing sampling intervals when using DMD for both the candidate satellites. This test was done by setting the number of time delays used to a constant value across the two satellites and varying the sampling interval while performing DMD. The SGP4 datasets for each satellite were used, for consistency. In the scenario shown, ISS used 20-time delays and MOLNIYA-3-50 used 110-time delays, a number selected arbitrarily to exceed the value identified in the previous section. Here, the DMD algorithm will find the reduced order matrix $\bsym{\Tilde{A}}$, through the singular value thresholding criterion. 
It is seen that $\Delta k = 7$ minutes and $\Delta k = 7$ minutes are ideal for the SGP4 datasets of ISS and MOLNIYA-3-50 satellites, respectively.

\section{Conclusion}
\label{Sec:Conc}
This work discussed the predictive capabilities of DMD and its suitability for application in the context of orbital mechanics. Theoretical development on the minimum number of time delays required to model a nonlinear periodic trajectory using an AR model was presented with connections to the DMD algorithm. The analysis is similar to the work found in ~\cite{pan2020structure}, but is reformulated concisely and structured for the problem considered here. This work shows that there always exists an AR model for a periodic orbit (of a nonlinear system) with a finite number of harmonics. Further, the minimal time delays required can be construed directly from the data matrix and is related to the number of dominant frequencies in the orbit. In addition, the use of Hankel-DMD to learn satellite trajectories subject to different perturbations was investigated. Methods to identify suitable training parameters such as the minimum number of time delays and training window size were discussed. Further, the accuracy of making predictions using the above parameters is also mentioned. In the future, the use of an extended-DMD algorithm with theoretical grounds for the use of suitable basis functions and time delays will be explored.

\section*{Acknowledgements}
The authors acknowledge with gratitude the help and support from Dr. Michael Yakes (Program Officer - Physics of Remote Sensing Program). This work is supported by the Air Force Office of Scientific Research under Grant No. FA9550-20-1-0083.


\bibliographystyle{asmejour}   

\bibliography{asmejour-sample} 

\appendix
\section{Two line element sets for ISS and MOLNIYA-3-50}
\label{App:A}

\begin{figure}[h]
    \centering
    \includegraphics[scale=0.5]{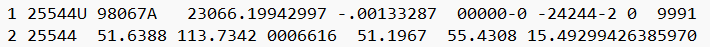}
    \caption{ISS Two-line element set}
    \label{fig:TLE_ISS}
\end{figure}

\begin{figure}[h]
    \centering
    \includegraphics[scale=0.5]{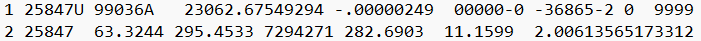}
    \caption{MOLNIYA-3-50 Two-line element set}
    \label{fig:TLE_ML-350}
\end{figure}

Figs.~\eqref{fig:TLE_ISS} and ~\eqref{fig:TLE_ML-350} show the two-line element considered in this work for the ISS and MOLNIYA-3-50 satellites. The corresponding earth orientation parameter and space weather data can be obtained from \href{http://www.celestrak.org/SpaceData/}{Celestrak}.

\section{Dataset Availability}
The datasets supporting the findings in this paper can be openly accessed/downloaded at \href{https://github.com/sriramn1122/DMD-Orbital-Datasets/tree/main}{DMD-Orbital data}



\end{document}